\documentclass[preprint]{aastex}

\newcommand{\calr}{\ifmmode{\cal R}\else$\cal R$\fi}
\newcommand{\stacs}{{\sc StaCS}}
\newcommand{\msun}{M_\odot}
\newcommand{\be}{\begin{equation}}
\newcommand{\ee}{\end{equation}}
\newcommand{\omegam}{\Omega_{\rm m}}
\newcommand{\sigmaeight}{\sigma_8}
\newcommand{\omegal}{\Omega_\Lambda}

\newcommand{\gtsima}{$ \buildrel > \over \sim \,$}
\newcommand{\ltsima}{$ \buildrel < \over \sim \,$}
\newcommand{\simgt}{\lower.5ex\hbox{\gtsima}}
\newcommand{\simlt}{\lower.5ex\hbox{\ltsima}}

\newcommand{\hmpc}{h^{-1}\,{\rm Mpc}}
\newcommand{\kms}{\ifmmode\,{\rm km}\,{\rm s}^{-1}\else km$\,$s$^{-1}$\fi}
\newcommand{\degs}{\ifmmode^\circ\else$^\circ$\fi}
\newcommand{\sqdegs}{\ifmmode\Box^\circ\else$\Box^\circ$\fi}
\newcommand{\kmsmpc}{\kms\,{{\rm Mpc}}^{-1}}
\newcommand{\etal}{et al.}
\newcommand{\clusterfind}{{\sc clusterfind}}
\newcommand{\lcoarse}{\mbox{${\cal L}_{\rm coarse}$}}
\newcommand{\lfine}{\mbox{${\cal L}_{\rm fine}$}}
\newcommand{\Lam}{\mbox{$\calr$}}
\newcommand{\mstar}{\mbox{$M^\ast$}}
\newcommand{\lstar}{\mbox{$L^\ast$}}
\newcommand{\nfield}{\mbox{$N_f$}}
\newcommand{\rcore}{\mbox{$r_{core}$}}
\newcommand{\rmax}{\mbox{$r_{max}$}}
\newcommand{\mlim}{\mbox{$m_{lim}$}}

\newcommand{\z}{\mbox{$z$}}

\shorttitle{STACS}
\shortauthors{Willick \etal}

\begin{document}


\title{The Stanford Cluster Search for Distant Galaxy Clusters
\footnote{The Hobby-Eberly Telescope (HET) is a joint project of
the University of Texas at Austin, the Pennsylvania State University,
Stanford University, Ludwig-Maximillians-Universit\"at M\"unchen,
and Georg-August-Universit\"at G\"ottingen.  The HET is named in
honor of its principal benefactors, William P. Hobby and Robert E.
Eberly.}}


\author{Jeffrey A.\ Willick\footnote{Deceased.}, Keith L.\ Thompson, \& Benjamin F.\ Mathiesen}
\affil{Department of Physics, Stanford University, 
    Stanford, CA 94305-4060;
    klt@perseus.stanford.edu, bfm@perseus.stanford.edu}
\author{Saul Perlmutter\footnote{Center for Particle Astrophysics, 
  University of California}, Robert A. Knop}
\affil{Institute for Nuclear and Particle Astrophysics,
    E. O. Lawrence Berkeley National Laboratory, Berkeley, CA 94720}
\and
\author{Gary J. Hill}
\affil{McDonald Observatory, University of Texas, Austin, TX 78712}

\begin{abstract}

We describe the scientific motivation behind, and
the methodology of,  the Stanford Cluster Search (\stacs), a
program to compile a catalog of optically selected clusters
of galaxies at intermediate and high
($0.3\,\simlt z \simlt\,1$) redshifts. The clusters
are identified using an matched filter algorithm  
applied to deep CCD images covering $\sim 60$ 
square degrees of sky. These images are obtained from
several data archives, principally that of the Berkeley
Supernova Cosmology Project of Perlmutter \etal\ 
Potential clusters are confirmed with spectroscopic 
observations at the 9.2 m Hobby-Eberly Telescope. Follow-up
observations at optical, sub-mm, and X-ray wavelengths are 
planned in order to estimate cluster masses. Our long-term
scientific goal is to measure the cluster number density
as a function of mass and redshift, $n(M,z)$, which
is sensitive to the cosmological density parameter $\omegam$
and the amplitude of density fluctuations $\sigmaeight.$ 
nd the amplitude of density
fluctuations on cluster scales.  Our short-term goals are the detection
of high-redshift cluster candidates over a broad mass range and the
measurement of evolution in cluster scaling relations.
The combined data set will contain clusters ranging over
an order of magnitude in mass, and allow constraints on
these parameters accurate to $\sim 10\%.$ We present our
first spectroscopically confirmed cluster candidates and
describe how to access them electronically. 

\end{abstract}





\section{Introduction}

The mechanisms and timescales for the formation of 
massive, gravitationally bound objects
are topics of considerable interest in cosmology.
It is likely that such objects arise via
the gradual growth of small ($\delta \equiv \delta\rho/\rho \ll 1$)
primordial density fluctuations, followed
by rapid gravitational collapse after $\delta \sim 1.$ In
hierarchical structure formation scenarios, 
the fluctuation amplitudes are greater on small scales
than on large ones;
thus, the formation epoch of gravitationally
bound objects increases with mass. In most scenarios,
galaxies (mass $M_g \sim 10^{11}$--$10^{12}\,\msun$) form
early, $z \approx 2$--$10,$ while rich galaxy clusters 
($M_c \sim 10^{14}$--$10^{15}\,\msun$) collapse more recently
($z \simlt 1$), and the most massive ones
may be still forming today.

About twenty-five years ago \citet{ps} developed
the formalism that has been used most widely to
quantify the above ideas. They found that 
the comoving number density of collapsed
objects of mass $M$ has the form
\be
n(M,z) \propto \nu_{_M} e^{-\nu_M^2/2} \,,
\label{ps0}
\ee
where $\nu_{_M} \approx \delta_c/\sigma_M(z),$ 
$\sigma_M(z)$ is the rms fractional mass fluctuation
on mass scale $M$ at redshift $z,$ 
and $\delta_c \approx 1.69 $ is a numerical constant that
is virtually independent of cosmology and epoch. 
We discuss the PS approach in greater detail in \S 2 below.
For now, note that exponential character of Eq.~(\ref{ps0})
implies a strong sensitivity of $n(M,z)$ to 
several cosmological factors: (1) the overall amplitude of 
mass fluctuations, parameterized, e.g., by its amplitude
on a fiducial scale, such as $\sigmaeight,$ the present rms fluctuation
on an $8\hmpc$ scale; (2) mass $M$ itself, as $\sigma_M$
is a rapidly decreasing function of mass; redshift
$z,$ because linear fluctuation growth causes $\sigma_M(z)$
to increase with cosmic time; and
finally, the density parameter $\omegam,$
which controls the epoch $z \sim \omegam^{-1}$ at which
the universe enters free expansion and structure freezes
out.\footnote{In a flat ($\omegam+\omegal=1$), low-density
universe, linear growth continues to a
later epoch than in an open universe of the same $\omegam.$}

As a consequence of the sensitive dependence of $n(M,z)$
on $\omegam$ and $\sigmaeight,$ an accurate measurement of
$n(M,z)$ can yield strong constraints on these parameters.
Although it is possible in principle to obtain such constraints
using objects of any mass, it is the rich clusters
of galaxies, with $M \simgt 3\times 10^{14} \msun,$ that are best
suited to the task. For one thing, 
they form at relatively low redshift and thus can be
identified and studied, with moderate observational
effort, at or near their formation epoch. To do the
same for galaxies is observationally
challenging at present. In addition, rich clusters
are susceptible to accurate virial mass estimates 
using a variety of techniques. For galaxies this is
more difficult because their virial radii 
are generally much larger than their visible extent.
As we clarify below, measurement of the virial mass
in the specific sense considered by PS is
crucial to a reliable application of the formalism.
Finally, clusters are massive enough that hydrodynamic
and radiative transfer effects are not expected to
play a significant role in their formation history,
whereas for galaxies they might, and it is only to the
degree that formation is purely gravitational that
the PS theory is valid.

As was first pointed out clearly by \citet{wef93}, even
a measurement of $n(M,0),$ the cluster mass function
at the present epoch, constrains $\omegam$ and $\sigmaeight,$ albeit
in a degenerate combination: $\sigmaeight \omegam^{0.5} \simeq
0.5$--$0.6.$ To break the degeneracy, one must probe
the cluster mass function at a range of redshifts
out to $z\,\simlt\,1.$ One then finds that, given the
$z=0$ constraint, the evolution of the cluster number
density is a strong function of $\omegam.$ For Einstein-de
Sitter universes, $n(M,z)$ drops precipitously with increasing redshift
for massive ($M \simgt 3\times 10^{14} \msun$)
clusters, while for low-density models ($\omegam \simlt 0.3$)
the decrease in cluster abundance with redshift is very
gradual. This basic fact has been emphasized, and tentative
cosmological constraints derived, by numerous
authors in recent years 
\citep{bfc97,ob92,ob97,
carl97,bf98,donahue98,
bb98,bbb98,sbo98,ecfh98, 
R99,bsbd99,vl99a,vl99b,bcdoy00,hmh00}.

Despite the promising nature of this cosmological
test, its goals have not yet been realized. The clearest
evidence for this is the discrepancies among the
conclusions drawn by the aforementioned authors. At this
risk of oversimplifying a bit, the results to date fall
into two ``camps.'' One, exemplified by Blanchard and
coworkers, finds strong evolution in the cluster abundance
with redshift, and consequently a high
value of the density parameter, $\omegam \sim 0.8 \pm 0.2$
\citep{bsbd99,R99,vl99a,vl99b}. The opposing
camp, associated principally with Bahcall and coworkers,
has argued that $n(M,z)$ decreases little, if at all,
out to $z \sim 0.8,$ and consequently favors a low
density universe, $\omegam \sim 0.2 \pm 0.1$ 
(\citealp{bcdoy00}; see also \citealp{carl97}, \citealp{donahue98}).  
Somewhere in between these extremes lies the work of
\citet{ecf96,ecfh98}
who find $\omegam \approx 0.45 \pm 0.2.$
Generally speaking, the Bahcall camp finds results consistent
with the low-density, flat universe favored by the combination
of CMB anisotropies and SN Ia observations 
\citep[e.g.][]{L00,b00},
while the Blanchard camp finds 
the cluster evolution to be consistent with a critical
density universe. 

The reasons for these rather different estimates of $\omegam$ 
are complex and
poorly undersood, and we will not attempt to do justice
to them here. The interested reader is referred to the
thoughtful discussion by \citet{ecfh98} (see in particular
their \S 5) for further insight. We note, however, that
the studies cited above all rely, in whole in part, on X-ray selected
samples of galaxies, and in particular, the Einstein Medium Sensitivity
Survey (EMSS) of rich clusters. These studies must thus make crucial
assmptions about the relationship between the X-ray properties
of clusters and their masses, and about the way these X-ray properties
may (or may not) evolve with redshift. In addition, in order to
convert a sample of X-ray detected clusters into an estimate
of $n(M,z),$ one must know how the selection function of clusters
depends on X-ray flux and redshift. It is not obvious
that the EMSS selection function is known, to the needed accuracy,
at the high redshifts and low fluxes that are crucial to
the cosmological measurement. Future X-ray catalogs derived
from ROSAT and supplemented by data from Chandra and XMM, or
based on XMM alone \citep{rvlm99}, are likely to ameliorate
these problems, but may not solve them completely.

A strong argument may thus be made for basing the
cluster abundance test on optically selected cluster catalogs. 
Such an approach may run afoul of a long-held view that
X-ray emission from the hot intracluster medium is a strong
indicator of true virialization, whereas optical overdensities
on the sky may result from superpositions of nonvirialized
objects 
\citep{fews96,vhfw97}.  
Such arguments, though valid in principle, do not exclude
the possibility of constructing an unbiased sample of clusters
from optical imaging data {\em provided the cluster candidates
are followed up with extensive redshift measurements to confirm
that they are indeed virialized structures.} Moreover,  
the construction of such catalogs has been facilitated in
recent years by the development of sophisticated
automated cluster identification algorithms that can be applied
to large imaging databases. Several approaches to this
problem are possible, but the most widely used is the {\em matched
filter algorithm} and its variants. Proposed intially by
\citet{post96}, this approach has been further elaborated
by \citet{k98}, \citet{sb98}, \citet{kep99}, and \citet{kk00}
and successfully applied to data sets such as the Palomar Distant
Cluster Survey and the Sloan Digital Sky Survey.

Another point in favor of optical surveys is that they
have a greater ability to find low-mass clusters at high
redshifts. Matched filter algorithms are very efficient
detectors of ``rich groups'', containing several tens of
galaxies in what appears to be a compact structure. The
algorithm's great sensitivity to such near-clusters implies
a high degree of completeness for true clusters over a broad
mass range, and therefore greater statistical leverage on the
cosmological mass function. Data on the relative frequency
of clusters as a function of apparent richness can be found in
the above citations, as well as \citet{holden00}.
X--ray surveys, in contrast, are generally flux-limited and at
high redshifts only detect the most massive clusters in the
population.

In view of the scientific importance of accurately measuring
$n(M,z),$ and of the possible drawbacks of using an X-ray selected
sample for this purpose, we had begun a long-term research program
to create a complete survey of rich and poor clusters out to
redshifts of 1 by searching optical imaging databases. Our program
is known as the ``Stanford Cluster Search,'' or \stacs.
The purpose of this paper is to describe our project in detail
and to present preliminary results. Our intention is to make our
candidate clusters available to the community soon after we have
confirmed them with follow-up spectroscopy, which we carry out
mainly at the 9.2 m Hobby-Eberly Telescope \citep{het98}
at McDonald Observatory using the Marcario low-resolution 
spectrometer \citep{hill98}.  

Due to the death of JAW (the princple investigator),
the future of \stacs\ is unclear.  At the time of submission
we are limiting the scope of the project to approximately one more
year of candidate identification and confirmation
with HET, with the goal of generating a limited and
incomplete confirmed high-redshift cluster catalog.  
Given uncertainties in the future funding and time
constraints of KLT and BFM, however, we have decided to
finish and submit this paper to describe the current state of
the project as if we expected to continue as originally planned.  
We beg the indulgence of the reader with respect to various
references to long-term, and perhaps unrealistic, science goals.
We will also describe the short-term goals which can be
achieved with existing data over the next year or two.

The outline of this paper is as follows.
In \S 2 we further discuss some of the theoretical issues
involved in using clusters as cosmological probes.
In \S 3 we describe the use of deep archival images to identify 
cluster candidates, including a review of the matched
filter algorithm.  
In \S 4 we present the steps in processing
one archival field and producing candidate clusters,
and then present HET spectroscopic confirmation of one.  
We also discuss some limitations apparent in the 
use of matched filter processing of images without
spectroscopic confirmation.  
In \S 5 we discuss how \stacs\ figures into the growing
number of distant cluster programs now planned or in progress,
and outline the issues we will address in a series of papers
to be submitted over the coming year.

\section{Theoretical Background}

As already noted, the theoretical basis for
using clusters to constrain cosmological parameters
is best stated using the Press-Schechter
formalism. N-body simulations show that the PS
predictions are quite accurate for predicting
the cluster abundance as a function of mass and redshift
\citep{vl99a,vl99b,borg99}.  
We emphasize that the PS approach
is not necessarily the final word, and that ongoing
numerical experiments will lead to more accurate
semi-analytic formulae for $n(M,z).$ 
We discuss recent developments in this area
at the end of this section.
However, while such advances may bring about changes in detail,
particularly at very high masses and redshifts, they are
unlikely to change the basic picture we now outline. 
 
The PS {\em ansatz\/}
is that virialized structures of mass $M$ form when growing
density fluctuations
reach a certain threshold overdensity. This overdensity,
$\delta_c \approx 1.69,$ is derived from a simple spherical
collapse model, and thus does not necessarily describe
an individual collapsed structure, but does
describe well the ensemble of virialized structures
at a particular epoch. 
Let $\sigma_M$ denote the rms overdensity
on a mass scale $M,$ {\em extrapolated using the
linear growth rate to $z=0.$}
The PS formalism then yields the following expression
for $n(M,z)\,dM$ the number of virialized objects per unit
comoving density with masses between $M$ and $M+dM:$
\be
n(M,z)\,dM \,=
\sqrt{\frac{2}{\pi}}\,\frac{\overline\rho}{M^2}\,
\frac{\delta_c(z)}{\sigma_M}\left|\frac{d\ln \sigma_M}{d\ln M}\right|
\,e^{-\delta_c(z)^2/2\sigma_M^2}\,dM\,,
\label{eq:PS}
\ee
where $\overline\rho\equiv\omegam\rho_{crit}$ is the comoving mean
mass density, and $\delta_c(z)=\delta_c/D(z),$ where $D(z)$
is the linear growth factor normalized to unity
at the present. (In the language of \S 1, $\delta_c/\sigma_M D(z) 
\equiv \nu_{_M}.$)

Eq.~(\ref{eq:PS}) is sensitive to the
background cosmology in
two ways. First, $D(z)$ depends on
the density parameters $\omegam$ (ordinary mass) and $\omegal$ (cosmological
constant or ``dark energy'').\footnote{The sensitivity of $D(z)$ to
$\omegal$ is considerably weaker than its sensitivity to $\omegam$
at the redshifts of interest, $z \simlt 1.$ Thus, the intermediate
redshift cluster abundance test cannot readily distinguish between
flat and open models of the same $\omegam.$ If the test can be
extended to redshifts $1 \simlt z \simlt 3,$ as proposed by
\citet{hmh00}, it can in principle be used to place strong
constraints on $\omegal$ as well.} 
In very low-density models ($\omegam \simlt 0.2$) 
structure formation virtually turns off at low redshifts
($z\simlt 1$), and thus the
$z\sim 1$ universe
differs relatively little from
the present in terms of the cluster abundance.
In contrast, critical-density universes form structure efficiently
into the present epoch, and many fewer massive
virialized objects are expected at $z \approx 1$ than are seen
at $z=0.$ Thus, all other things being equal, {\em the less
abundant massive clusters are in the past, 
the larger the density parameter must be.}
This is the essence of the cluster abundance test.

Eq.~(\ref{eq:PS}) also depends on cosmology through 
the rms mass fluctuation $\sigma_M,$ which
in turn is related to the primordial fluctuation spectrum
$P(k),$ one of the fundamental predictions of early-universe
theory:
\begin{equation}
\sigma_M^2 = \int_0^\infty \frac{dk}{k} \Delta^2(k)\,W^2(kR)\,,
\label{eq:sigmr}
\end{equation}
where $\Delta^2(k)\equiv k^3 P(k)/2\pi^2,$ and $W(kR)$ is the
``window function'' which picks out the spatial scale $R$
corresponding to the mass scale $M.$ The shape of the power spectrum $P(k)$
is determined, in the Cold Dark Matter (CDM) class of cosmological
models, mainly by the cosmological parameters $\omegam$ and $H_0.$ 
Its amplitude is constrained, for given values
of the parameters $\omegam,$ $\omegal,$ and
$H_0,$ by the requirement that the predicted
large-angle anisotropies in the 
Cosmic Microwave Background (CMB) radiation match those
observed by the COBE satellite. (Alternatively, the amplitude
of $P(k)$ can be constrained by the abundance of clusters
at the present epoch; see below.)

The specifically Gaussian form of Eq.~(\ref{eq:PS})
stems from the assumption that initial mass fluctuations are Gaussian
in nature, an assumption that follows naturally from
inflationary scenarios, and which has yet to be contradicted
by observation (\citet{bhl00}). It is important to note that
the Gaussian factor results in
a minuscule space density of extremely high mass
(small $\sigma_M$) clusters, especially at redshifts
approaching unity. Thus, the presence of such clusters
could be an indication of nongaussianity in the initial
fluctuation spectrum. See \citet{w00} for the details of,
and caveats about, this argument. Better data on the
abundance of very massive clusters at $z \sim 1$
can thus shed light on the Gaussianity (or lack
thereof) of the primordial
mass fluctuations, an important ancillary goal
of the cluster abundance test.

\subsection{Constraining $\omegam$}

Assuming the primordial fluctuations are indeed Gaussian,
the cluster abundance at intermediate redshifts depends almost
exclusively on $\omegam.$ 
So strong is the dependence
on $\omegam$ that even modest changes in the density
parameter can substantially
change the predicted number of clusters on the sky. 
To illustrate this, we use Eq.(\ref{eq:PS}) to
compute the number of clusters per unit solid angle,
as a function of mass, for three values of the density
parameter, $\omegam=0.25,$ $0.35,$ and $1.00,$  
assuming a flat universe. The sky density of clusters
more massive than a given threshold $M,$ and lying
at redshifts $\le 1$
(the effective limiting redshift we expect for \stacs)
is given by
\begin{equation}
N(\ge M,\le z) = \int_0^1 dz\,\frac{dV}{dz}
\int_M^\infty dM\, 
n(M,z) \,,
\label{ngtmz}
\end{equation}
where $dV/dz$ is the comoving volume per unit redshift.

The results of this calculation are shown in the left 
panel of 
Figure~1,
where a survey area
of 25 square degrees---the minimum we expect of \stacs\
in its initial phase---is assumed. We see that
the predicted sky density of clusters is vastly smaller
in the critical density model than for the two low density models.
Indeed, only a handful of rich ($M\simgt 3\times 10^{14}\,\msun$
clusters are expected in the \stacs\ minimal survey
area if $\omegam=1.$ 
By contrast, several tens of clusters are anticipated in
the low density models. This illustrates the power
of the test to distinguish low-density from Einstein de-Sitter
models.

\begin{figure}[t!]
\centering
\includegraphics[scale=0.40]{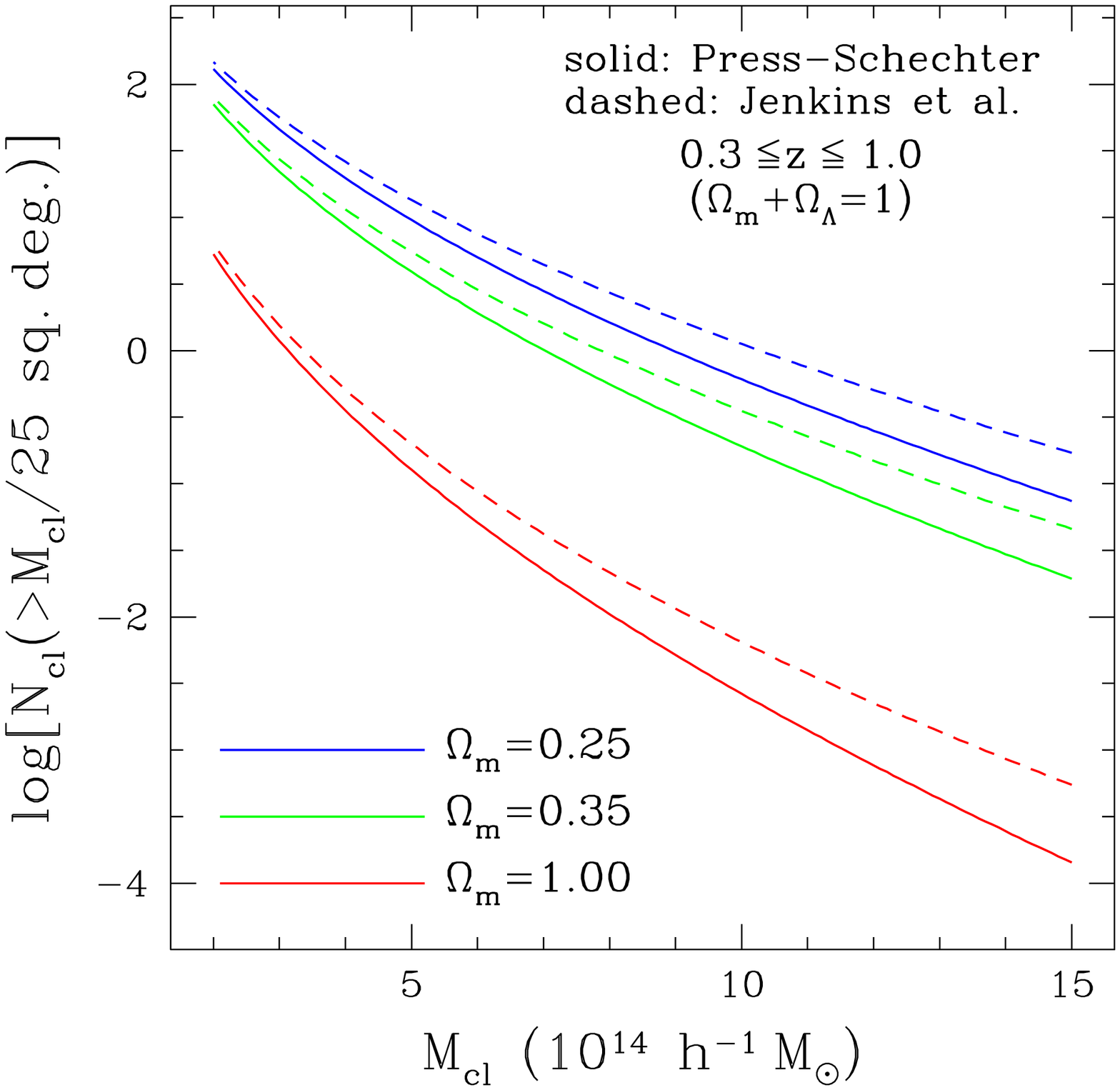}\hfill~
\includegraphics[scale=0.40]{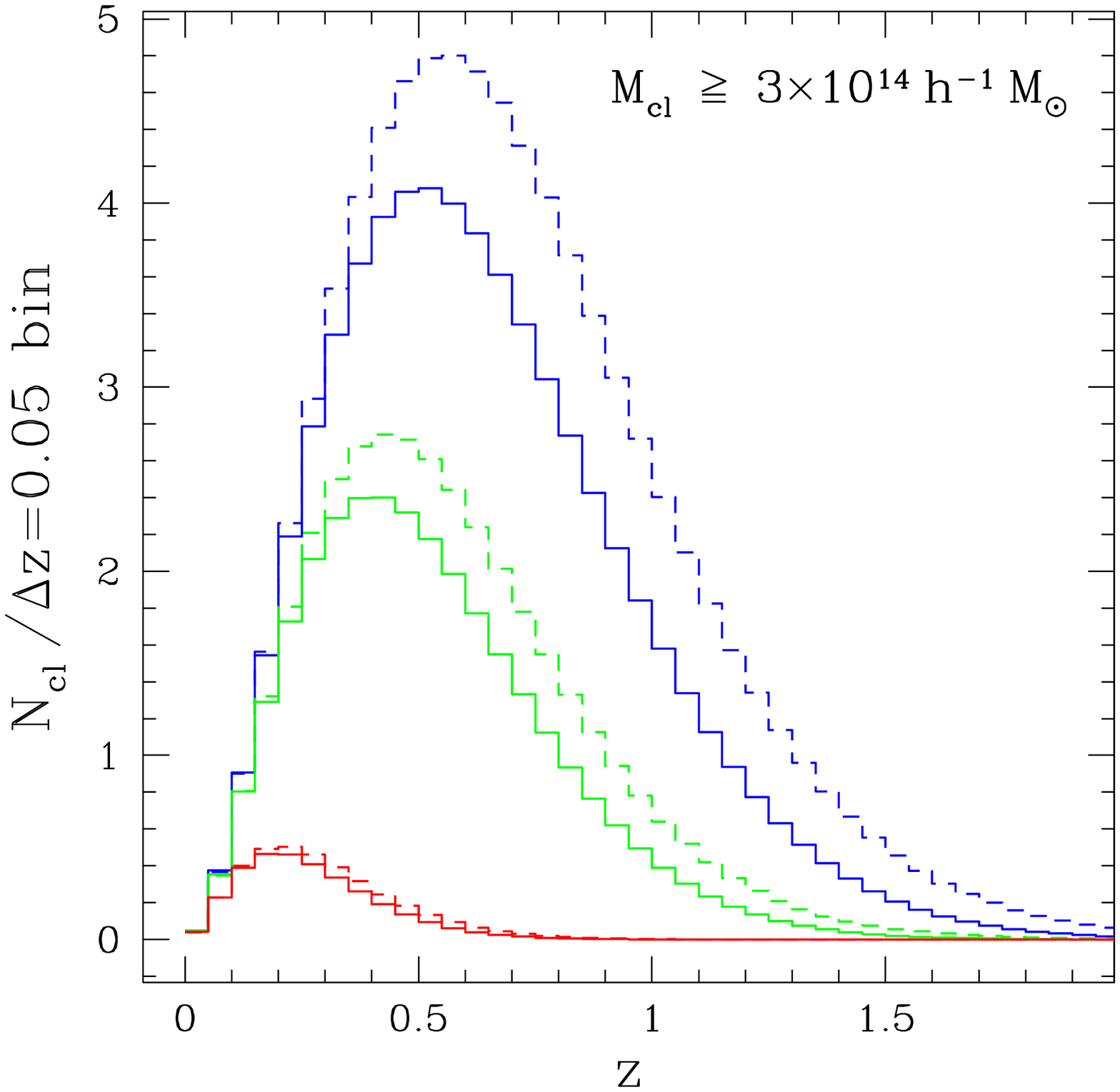}
\caption{{\small The expected number of clusters per square degree of
sky, with redshifts $z \le 1,$ and mass $\ge M,$
for two representative low-density, flat cosmologies. The
calculations are made using the Press-Schechter formalism. 
The rms mass fluctuations are
normalized to the present-day cluster abundance.}}
\label{fig:abund}
\end{figure}

The right panel of
Figure~\ref{fig:abund}
shows the redshift
dependence of the cluster abundance for the different values
of $\omegam.$ Presented this way, one sees that even the two
low-density models differ substantially; at the higher redshifts
$z \simgt 0.5$ the $\omegam=0.25$ model yields 2--3 times as many
rich clusters per unit redshift as the $\omegam=0.35$ model. This
extreme sensitivy to $\omegam$ 
is the basis of our ability to constrain $\omegam$ rather
precisely, to a degree we now quantify.

The right panel of
Figure~\ref{fig:abund}
shows a dramatic
decrease in the sky density of clusters with mass $M_{\rm lim}$
$\ge 3\times 10^{14} \msun$ as $\omegam$ increases.
Integrating the histograms out to $z = 1$, we find a total
cluster count of 2.15 per square degree for the $\omegam = 0.25$
cosmology and 1.14 per square degree for the $\omegam = 0.35$
cosmology. We can therefore calculate that
$d\ln[N(\ge M_{\rm lim},z\le 1)]/d\ln\omegam = -1.88$ for our particular
choice of cosmological parameters in the Press-Schechter formalism.
It follows that $|\delta\omegam/\omegam| = 0.53\delta N/N.$
Assuming Poisson statistics, a sky density appropriate for
$\omegam = 0.3$, and 25 square degrees in the 
Supernova Cosmology Project (SCP) data base, we would
expect to obtain about 41 clusters above our threshold
mass for $\delta N/N = 0.16$
and $\delta\omegam/\omegam = 0.08$.  It is therefore 
reasonable to expect $\sim 10\%$ precision in $\omegam$ from
a complete survey of just the SCP imaging database. This
perhaps surprising figure arises from the number of clusters
detectable by optical surveys, which is much greater than
that obtainable through X--rays at high redshifts.

As discussed in \S 3, by increasing our sky coverage
with other imaging databases, we can further improve
these constraints. The limiting factors on our
uncertainty in $\omegam$ are the number of clusters
found, our ability to evaluate the survey completeness
function, the accuracy of our mass function model,
and our ability to set a limiting mass threshold.
The problem of measuring cluster masses accurately is common to
all cluster surveys, and requires extensive follow-up
observations.  Redshifts obtained through the Hobby-Eberly
telescope will help a great deal in this regard, but
much more accurate masses can be obtained through weak
lensing, X--ray spectroscopy, or combined analysis of
X--ray and Sunyaev-Zel'dovich surface brightness maps.
\stacs is currently pursuing all three options for 
follow-up of confirmed clusters.
 
\subsection{The Importance of the Low-Redshift Constraint}
An important aspect of the calculation depicted in
Figure~\ref{fig:abund},
alluded to but not fully explained
above, warrants further comment.
When the PS formalism is applied to
the {\em present-day density of rich clusters,} which is
thought to be well known, a degenerate constraint on
$\sigma_8$ and $\omegam$ is obtained. The constraint
used in the calculation above, obtained by \citet{borg99}
for a flat universe, is
\begin{equation}
\sigma_8 = (0.58\pm 0.06) \times \omegam^{-0.47+0.16\omegam}\,.
\label{eq:B99}
\end{equation}
Equation~\ref{eq:B99} allows the amplitude
of the fluctuation spectrum $P(k)$ to be determined for
any chosen $\omegam,$ and thus allows $\sigma_M$ to be
calculated for any cluster mass $M.$ By implication, our
prediction of the number of intermediate redshift clusters
is only as good as the low-$z$ normalization. There is
now some reason to doubt it, though, as recent work
\citep{post96,holden99,holden00} 
has found tentative evidence that the low-redshift cluster
abundance is considerably higher than indicated by
the Abell Cluster Catalog 
\citep{abell89},
long thought to be nearly complete to $z=0.2.$ 
It is not yet clear what these results
may imply for the accepted normalizations such as 
Equation~\ref{eq:B99}. They do, however, suggest that 
\stacs\ and other cluster search programs cannot necessarily
assume a known low-$z$ normalization of $\sigma_8,$
but rather must determine it self-consistently
along with $\omegam$ from the higher redshift cluster abundance data.

\subsection{Determination of Virial Masses}
\label{sec:clmass}

An oft-neglected subtlety in applying the PS formalism
is the precise definition of {\em virial mass,} the 
mass that properly is to be used in Eq.~(\ref{eq:PS}).
Gravitationally bound objects are not, in general,
truncated abruptly at a particular radius; instead, their
density profiles $\rho(r)$ smoothly approach the background mean density
$\overline{\rho(z)}.$ Hence, one cannot specify their mass without
reference to a particular radius or overdensity. 
Eq.~(\ref{eq:PS}) applies specifically to the mass $M_V$ within
a radius, $r_V,$ with the property
$3 M_V/4\pi r_V^3 = \Delta_V(z,\omegam,\omegal)
\overline{\rho (z)}.$ 
In an $\omegam=1$ universe, $\Delta_V\simeq 178$
at all redshifts; for $\omegam<1,$ $\Delta_V$ is larger
than this value, and increases with decreasing redshift.
\citet{ks96} give analytic
approximations to $\Delta_V(z)$ as a function of
$\omegam$ for flat and open cosmologies.

This definition of virial mass means
that one cannot compute $M_V$ from
observational data, such as galaxy velocity
dispersion or X-ray temperature, without specifying
both a cosmology ($\omegam$ and $\omegal$) and a
model for the radial mass distribution of the cluster.
Willick (2000; hereafter W00) studied this issue
for MS1054-03, a massive cluster at $z=0.83.$
He showed that one could convert velocity dispersion,
X-ray temperature, and weak lensing data for this cluster
to an accurate virial mass if (1) a mass profile of, e.g., the type advocated
by \citet{nfw} (NFW) was adopted (though any
other simple analytical model would have worked just as well), and (2)
the characteristic radius $r_s$ for this profile was
taken as known.
If no density profile was assumed a priori,
the virial mass was determined only to within about 50\%;
even given the choice of the NFW profile
the uncertainty was at least 20\% (not counting
observational uncertainties) without adopting
a particular value for $r_s.$ 
The latter uncertainty is illustrated in Figure 4 of W00.  
W00 argued that 
preferred values of $r_s,$ and thus of $M_V,$ could be chosen by requiring
that the cluster concentration index (essentially the
ratio of $r_V$ to $r_s$) be in the relatively narrow
range predicted from the N-body experiments of NFW.
However, this constraint is, in truth, based on theoretical
analyses not yet confirmed by detailed observational study.

W00's analysis reinforced an issue that
should be taken very seriously in future cluster analyses:
{\em The density profiles of clusters are crucial to
determining the virial masses which enter into the PS formalism.}
And yet, we still are not certain whether the NFW profile,
or any other analytic form, for that matter, are valid for
a majority of rich clusters. This question relates to
the broader and deeper issue of, What is the distribution
of dark matter in gravitationally bound structures? 
Any analysis of clusters should
pay as much attention to this question as to that of
the evolution of the their abundance over cosmic times.

This is, then, a secondary goal of \stacs. The
clusters we identify will be followed up by
observations that aim to constrain cluster density profiles
as well as measure mass. The most useful data for this
purpose will be weak gravitational lensing data, which
enable one to compute the run of cluster mass with radius
out to the virial radius ($r_V \approx 1.5\hmpc$) and beyond.
(By contrast, strong gravitational lensing, such as giant
arcs, typically constrain the mass only within a few
hundred kiloparsecs of the cluster center.) 
Such data are best obtained from space telescopes
such as HST and, eventually, NGST, although the new generation
of ground-based telescopes such as  VLT and Gemini, with their
much larger fields of view and
promise of image quality approaching that of the HST, may
end up being as well or better suited to this purpose.
Galaxy velocity data and X-ray data obtained using Chandra can
also shed light on this problem, and we will pursue these
avenues as well. By publishing our \stacs\ clusters following
spectroscopic confirmation, we hope that community follow-up
will lead to the acquisition of such data. The long-term
goal of accurately measuring the mass distribution in
rich clusters is of necessity the task of many scientists
pooling observational resources. 

\subsection{A Breakdown of Press-Schechter?}

In important issue for \stacs, and indeed for any
program that uses clusters as cosmological probes, is
the possible breakdown of the PS approach at 
masses and epochs such that $\nu_M = \delta_c(z)/\sigma_M \gg 1.$
The PS formalism suggests
that such objects are exponentially rare, with abundance
$\sim \nu_M e^{-\nu_M^2/2}.$ As a result, the existence
of very massive ($M \ga 10^{15}\,\msun$) clusters at
redshifts approaching unity, such as the X-ray
clusters MS1054--03, has been taken as prima
facie evidence for low $\omegam$ \citep{bf98,donahue98} or
as possible evidence for non-Gaussian initial conditions
(W00).

Only recently have there been cosmological N-body simulations of
large enough volumes to test PS in this regime.
While these quite recent results
should be considered preliminary, it now appears that
the PS formula, Eq.~(\ref{eq:PS}), does indeed break down 
in the sense that it {\em underpredicts\/} the abundance
of the most massive clusters, especially at moderately
high ($z \sim 1$) redshift 
\citep{g98,g99,j00}.
This departure from the PS predictions means that quantitative
conclusions drawn from a small number of massive, high-redshift
clusters, such as those discussed in the previous paragraph,
cannot be correct in detail. 

The failure of Press-Schechter for the rarest objects
does not mean that clusters cannot be used as cosmological
probes. For one thing, recent analytic work to improve
upon PS has led to alternative formulae
that more accurately predict the cluster abundance
at the high-mass end \citep{smt99,st99}. 
It these formulae are confirmed, one can derive cosmological
constraints from cluster abundance data 
using the new formulae rather than the PS expression. 
For another, it still appears that the PS formalism
performs well in the intermediate cluster mass regime
($M \approx 3$--$5 \times 10^{14}\,\msun$), where
the majority of intermediate redshift clusters are found
in any case. 

Thus, the cosmological tests we propose to do with \stacs, and
that are planned with other cluster surveys, remain viable.
However, we and other workers must take great care,
in the analysis phase, to account for departures from
the familiar Press-Schechter formalism that now, apparently,
occur at the high-mass end. New theoretical developments in 
this subject are
certain to emerge in the coming years, and we will follow them closely.

\section{Image Analysis and Cluster Identification}

We begin with existing archival imaging databases that cover
a significant area ($\simgt 10\,\sqdegs$) of the sky, and are complete
to a limiting magnitude of $m_R \simeq 23.5.$ The most important
imaging database we use is that obtained by the Supernova
Cosmology Project \citep[SCP;][]{perl99}.  
The SCP images
were obtained as part of a search for distant supernovae,
but are suited to identifying intermediate
redshift clusters as well. The SCP imaging database currently
covers approximately 25 square degrees of sky (see below). 
We are also working, in collaboration
with M.\ Postman, with images
from the DEEPRANGE survey \citep{deeprange}, which covers 16 square
degrees of sky, and with images from the ESO Imaging Survey 
(\citealp{nonino99}
\footnote{For further information see the EIS Web 
page, {\tt http://www.eso.org/science/eis}.}),
which covers 24 square degrees of  sky. This total database of
over 60 square degrees to $m_{R,lim} \simgt 23.5$ will enable
us to detect many ($\sim 50$--100) rich clusters, with nearly
100\% detection effeciency, in the redshift range $z \simeq 0.3$--$1.0$.

\subsection{Data Reduction Procedures}

In this section we describe the steps required to go from
raw CCD images, acquired from the archival imaging databases
described in \S 3, to catalog of cluster candidates with
estimated masses and redshifts. We use a variety of publically
available and commercial data-processing software in this process, including
IRAF\footnote{IRAF is distributed by the
National Optical Astronomy Observatories,
which are operated by the Association of Universities for Research
in Astronomy, Inc., under cooperative agreement with the
National Science Foundation.}, STSDAS, FOCAS \citep{jt81},  
SEXTRACTOR, and IDL, plus locally written software. 

\subsubsection{Galaxy catalog generation from deep images}

The images we obtain from the SCP or EIS archives are already
flatfielded and sky-subtracted. Thus, the first stage in the
image reduction is stacking and coadding
several frames to reach the required limiting magnitude
$m_{R} \sim 23.5$ required for completeness to $z \simeq 1.$
Typically 4 or more frames with small dithers are coadded.
Registration and coadding are accomplished using the IRAF
package IMMATCHX. 
We have found that the distortions in a typical prime focus image are
large enough that generating mosaics is impractical, so we have used
coadded frames from sets of images shifted no more than
a few tens of arcseconds.  Given the nature of the SCP data base, the
images were taken over more than one observing run, separated in time
by a month or more. 

Following the generation of coadded frames, we run a sequence of 
FOCAS tasks that identify and categorize objects.  
FOCAS produces a list of galaxies and stars, with positions
in CCD coordinates and preliminary magnitudes. We then employ
locally written software to carry out astrometry and accurate
photometry. 
We use the USNO-A2.0 astrometric
catalog\footnote{The catalog is available electronically
from the US Naval Observatory's Flagstaff Station at
{\tt http://www.nofs.navy.mil.} See \citet{m98} for further information.}
which 30--150 unsaturated objects per frame to tie each coadded CCD
frame to an J2000 astrometric system. Typical rms positional errors  
are $0\farcs 5.$ Our photometry package improves upon FOCAS
in several ways:
\begin{enumerate} 
\item We have forced the FOCAS routines to adopt 
 the initial sky subtraction carried out
 by the SCP for object identification and moment calculations.  
 For our own photometry routine, we fit a planar sky to the
 frame, since the initial SCP sky subtraction 
 was globally accurate only to $\sim 0.2\sigma_{sky}$.  The planar fit
 removes small, large-scale gradients in the residual sky.
 We also compute a local sky for each object in the catalog,
 using it in favor of the global fitted sky unless there are
 too few pixels or the local sky shows too much variance.
\item All stars on the frame are masked prior to galaxy photometry,
 and any portion of a galaxy light profile in a masked region
 is properly compensated for with a symmetrically located pixel
 in the galaxy profile.  
\item We carry out photometry within a series of
 elliptical apertures, stopping when the surface brightness
 along an elliptical contour drops below the sky error. The
 aperture photometry is then extrapolated to a total magnitude
 using the method of moments described by \citet{w99}.
\item A useful step, we have found, is to flag and correct certain objects
 that have been erroneously classified as galaxies by FOCAS.
 We plot total magnitude versus effective
 surface brightness for all objects on a frame. Stars fall on a
 locus defined by a tight $m$--SB relation of slope unity,
 and at $R \lesssim 20.5$ fall on a well-defined locus with a 
 higher surface brightness than galaxies.  
 We reject those stars from the galaxy catalog
 produced by FOCAS.  
 Close binary stars and bright stars near (but below) the saturation limit
 both can be eliminated from the galaxy lists in this manner.  
 For $R > 20.5$, the star
 and galaxy distributions are merged, but the galaxies outnumber
 stars by a large factor and we have kept the FOCAS
 star/galaxy classifications \citep[cf.][]{k98}.  
\item Saturated stars, bright stars off the edge of the frame,
 and the increased noise near the edge due to dithering
 cause both spurious and unidentified objects.  We 
 catalog these regions by creating (by hand) a set of exclusion boxes
 for each frame, including an indication of how close to
 the edge reliable galaxy identifications are found.  
 While our algorithm for chosing these exclusion regions is
 somewhat qualitative, we have found the procedure promotes
 consistency and eliminates some spurious features in the 
 galaxy luminosity functions and the cluster likelihood
 maps (\lfine\, defined in \S 3.2.).  
\end{enumerate}
Our final galaxy catalog for a given frame thus consists
of CCD $(x,y)$ coordinates; RA and DEC accurate to $0\farcs 5$ 
for brighter galaxies, $m_R\simlt 21,$ and to $\sim 1\farcs 0$ for
fainter objects;
total magnitude and its estimated error; and several
auxilliary pieces of information: effective surface brightness
and radius, position angle, and ellipticity. These auxiliary data
are not used in the cluster finding procedure (see below) but
may be of interest at a later time.

For both the SCP and EIS databases, we generally
reduce multiple ($\sim 5$--$20$), slightly overlapping frames, which we
then combine into a single catalog covering from $0.2$--1 square degree.  
When combining we attempt to ensure that
a given galaxy appears only once in the final catalog. This
is important, as 
multiple appearances of single galaxies in the overlap
regions can produce a spurious clustering signal. 
The positional and photometric parameters of the
final catalog objects are obtained by averaging
the corresponding parameters of the galaxies
that appear in more than one input catalog.  
Objects found on different input frames with
coordinates that differ by $\le 2\farcs 0$ are judged
to be the same galaxy and are combined. This nearly
always yields the desired elimination of duplicates,
though on occasion we find faint, low-surface brightness
objects near the frame edges that appear twice in the final
catalog. However, the numbers involved
are small and the task of automatically identifying these cases 
difficult so we have not attempted a more sophisticated algorithm.  

    During the combination of multiple frames' galaxy catalogs
we also remove objects within each frame's exclusion boxes.  A 
set of exclusion boxes for the final catalog consists of the
logical union  of box area associated with each individual frame
followed by the combination between frames with a logical intersection.  

\subsection{Application of the Matched Filter Algorithm}

We search the galaxy catalogs for candidate clusters using 
a matched filter algorithm (MFA). Our implementation
of the MFA,
\clusterfind, closely follows the methods
of \citet{kep99}, except where specifically noted below.
Readers are referred to \citet{post96}, \citet{k98}, 
\citet{kep99}, and \cite{kk00}
for more extensive discussions of the MFA, but for the sake
of completeness we
present an outline of the approach below.

\subsubsection{Models of the Field and Cluster Galaxy Distributions}

The MFA approach begins with a model for the number of galaxies
at angular distance $\theta$ to within
$d\theta$ of a cluster center, and having flux $l$
to within $dl:$ 
\be
n(\theta,l;z)\,2\pi\,\theta\, d\theta dl =
\left[n_f(l) + \calr n_c(\theta,l;z)\right]\,
2\pi\,\theta\,d\theta\, dl\,.
\label{eq:model}
\ee
Here, $n_f(l)$ is the background or ``field'' galaxy
contribution, which is assumed to be spatially uniform (but see below), 
and $\calr n_c(\theta,l;z)$ is the cluster contribution.
The factor $\calr$ represents the overall ``richness''
of the cluster, to be defined more precisely below. 

Aside from the richness factor, all clusters at a given $z$ are assumed
to be identical. That is, $n_c(\theta,l;z)$ is a universal
function, which we take to be a truncated Plummer law
in space multiplied by a \citet{s76} luminosity function. 
Following \citet{kep99} we require that $n_c$ be normalized 
in the sense that 
\be
2\pi \int \theta d\theta \int dl \, n_c(\theta,l;z) L(l,z) = L^*\,,
\label{eq:normal}
\ee
where $L^*$ is the characteristic luminosity in the Schechter function.
With this definition, it follows that the richness $\calr$ is
the cluster luminosity in units of $L^*.$ Imposing this normalization
we find that $n_c$ is given by
\be
n_c(\theta,l;z) = \frac{\left[\left(1+\frac{\theta^2}{\theta_c^2}\right)^{-1}
- \left(1+y^2\right)^{-1}\right]\left(\frac{l}{l^*}\right)^{-\alpha}
e^{-l/l^*}}
{\pi\theta_c^2 l^* \left[\ln(1+y^2)-\frac{y}{1+y^2}\right]
\Gamma(2-\alpha)}\,,
\ee
where:
\begin{itemize}
\item[] $\alpha $ is the faint end slope of the
Schechter luminosity function;  
\item[] $y\equiv r_{max}/r_c,$ where $r_c$ is the cluster core radius
and $r_{max}$ is the truncation radius;
\item[] $\theta_c = r_c/d_A(z),$ where $d_A(z)$ is angular
diameter distance;
\item[] $l^*=L^*/4\pi d_L^2,$ where $d_L$ is luminosity distance
(in practice, $l^*$ is further corrected for K-correction and 
Galactic extinction
as discussed below); 
\item[] and $\Gamma(x) = \int_0^\infty t^{x-1} e^{-t} dt$ is
the usual Gamma-function.
\end{itemize}
We discuss specfic choices of $\alpha,$ $L^*,$ $r_c,$ $r_{max},$
$d_A,$ and $d_L$ in the next section.
 
For the field galaxy distribution we adopt a power-law model,
\be
n_f(l) = l_f^{-1} \left(\frac{l}{l_f}\right)^{-\beta} \frac{N_f}{\theta_f^2}\,.
\label{eq:fielddist}
\ee
For a Euclidean universe one expects $\beta=2.5;$ in practice
we find $\beta \approx 2.0$ 
in the magnitude range of interest. 
The quantities $l_f,$ $N_f,$ and $\theta_f$ are not 
independent, and in practice, we fix $\theta_f=1\farcm 0$  
and $l_f$ to correspond to $m_R=20.$ Then, only $\beta$
and $N_f$ remain to be determined, which we do 
on a catalog-by-catalog basis, as we explain further in \S 5.

\subsubsection{Likelihood Maximization}

Detecting clusters with the MFA entails 
maximizing the likelihood, given the above models for the cluster and field
galaxy distributions, that a cluster lies at a given
sky position and redshift. This likelihood is calculated  
on a grid of trial positions, $(\alpha_i,\delta_j,z_k),$
where $\alpha$ and $\delta$ are RA and DEC in decimal units
and $z$ is redshift.
The spatial grid $(i,j)$ runs over the portion of sky 
covered by the galaxy catalog with some desired resolution, 
and the redshift grid $k$ over the range of redshifts
at which one expects to find clusters. We let $z_k=0.2,0.3,...,1.0,$
which provides sufficient redshift discrimination.
At redshifts greater than 1, $l^*$ falls below our
detection threshold and the completeness of the matched
filter algorithm falls off rapidly.  At redshifts less
than 0.2, the survey volume is small and no new
detections are expected.

\citet{kep99} and \citet{kk00} show that one can detect clusters
by maximizing either ``coarse'' or ``fine'' likelihoods. The
former are less accurate, but are less computation-intensive
to calculate. \citet{kep99} suggest
that the coarse likelihood be used for a first pass through
the data set, and the fine likelihood used to refine
the calculation for candidate clusters detected in the first pass.
However, our catalogs are considerably smaller than those
anticipated by \citet{kep99}, and thus
we use only the fine likelihood. 
\clusterfind\ then requires about 4 hours of CPU
time per square degree of catalog searched on a Sparc Ultra 5
with a 270 MHz processor, a modest computational cost. 

The fine likelihood $\lfine$ is the log of the Poisson probability that
galaxies are found at their observed locations given the presence
of a cluster of richness $\calr$ at position $(\alpha,\delta,z)$
(cf. \citet{kep99} who define \lfine\ as the Poisson probability and
work with log(\lfine)). 
It is given by
\be
\lfine = \left[ 
\sum_{i=1}^N \ln \left[n_f(l_i) + \calr n_c(\theta_i,l_i;\alpha,\delta,z)\right] \right] - N_e\,, 
\label{eq:lfine}
\ee
\citep{kep99}, where the sum runs over all galaxies in the catalog,
and $N_e$ is the total expected number
of galaxies in the data set,
\begin{equation}
N_e =  2\pi\!\int \theta\,d\theta \int_{l_{min}}^\infty (n_f + \Lam n_c)\,dl
\label{eq:defNe}
\end{equation}
where $l_{min}$ is the flux limit of the catalog. If, following
\citet{kep99}, we now define
\be
\delta_i = \frac{n_c(\theta_i,l_i)}{n_f(l_i)}\,,
\label{eq:deltai}
\ee
we find after some algebra a simpler expression for the fine likelihood:
\be
\lfine = \left[ \sum_{\theta_i \le \theta_{max}} \ln (1+\calr \delta_i)\right]
- \calr \eta\,,
\label{eq:lfnew}
\ee
where
\begin{equation}
\eta \equiv \frac{\gamma(1-\alpha, l_{min}/l_*)}{\Gamma(2-\alpha)}
\label{eq:defeta}
\end{equation}
where $\gamma$ and $\Gamma$ are the usual incomplete and complete
Gamma-functions.\footnote{The expressions \ref{eq:lfine} and
\ref{eq:lfnew} are not equivalent. However, maximizing the first
with respect to richness at a given trial position and redshift
yields the same equation for richness as maximizing the second.
Thus, it is sufficient to use \ref{eq:lfnew}, a quantity defined
by a much smaller sum than the first.} (The specific form of
the quantity $\eta$ results from integrating the Schechter
function from $l_{min}$ to infinity, giving the relative number
of expected galaxies in the sample. There is nothing fundamental
about Eq.~(\ref{eq:defeta}) to the MFA algorithm.)

We have modified Eq. \ref{eq:lfnew} to accomodate the
exclusion boxes associated
with our real galaxy catalogs.  For each point in RA,DEC,$z$ space,
the second term  on the right-hand side ($\calr\eta$) is multiplied
by an additional factor $\epsilon^{-1}$ such that $\epsilon$ is the
spatial integral over the assumed normalized spatial distribution
(truncated Plummer law) within $\theta_{max}$ that is external
to the exclusion boxes.  
This extra factor therefore corrects for areas in which galaxies
could not be detected even if they were present.  
This exclusion box correction eliminated a great deal of spurious
signal in the $\lfine$ map near frame edges and saturated stars,
increasing the effective area of the survey by about 10\% and allowing
for a more accurate determination of candidate parameters near the
affected areas.

At a given grid position $(\alpha,\delta,z),$ we maximize
likelihood by setting $\partial\lfine/\partial\calr = 0.$ This
leads to the equation
\be
\sum_{\theta_i \le \theta_{max}} \frac{\delta_i}{1+\calr\delta_i} -
\eta = 0\,.
\label{eq:solveforlambda}
\ee
The value of $\calr$ which solves this equation is then
substituted back into Eq.~(\ref{eq:lfnew}) to obtain the
final value of $\lfine$ at that position. As noted by
\citet{kk00}, Eq.~(\ref{eq:solveforlambda}) requires a
numerical solution, which can be fairly time consuming,
especially as it must be done at each grid point. Indeed,
this is one reason that \citet{kep99} advocate that
the coarse likelihood be used first. We have written
code to solve Eq.~(\ref{eq:solveforlambda}) that uses
the Numerical Recipes \citep{press86}
algorithm ZBRENT, and is sufficiently rapid so as not to be
a major bottleneck in \clusterfind. 

\subsubsection{Using \lfine\ to identify clusters}

The procedure above produces three three-dimensional maps,
$\lfine(\alpha_i,\delta_j,z_k)$, $\calr(\alpha_i,\delta_j,z_k)$,
and $\epsilon(\alpha_i,\delta_j,z_k)$.  
Cluster candidates are found by searching the \lfine\
map for local maxima that satisfy the following
conditions.  First, we consider
only \lfine\ values where the corresponding richness value
is positive and where the exclusion boxes correction 
factor $\epsilon > 0.5$.  
A deficit of galaxies in a region of the catalog will
generate a high value of \lfine\ with a negative richness;
adoption of $\epsilon^{-1}$ into Eq.(\ref{eq:lfnew}) 
largely corrects this effect for deficits caused by 
data artifacts (and indicates where missing data is
excessive), but there are nevertheless true underdense
regions caused by large scale structure.  
Second, the 95th percentile \lfine\ value is found
for each \z\ plane of the \lfine\ map 
(the $\lfine$ value in grid points with $\calr < 0$ is set to
zero beforehand)
and used as a threshold. 
We consider only local maxima at or above this significance level.
Finally, we require that a local maximum be found,
at or above the 95\% significance level, at the same
RA and DEC for three successive trial redshifts, e.g.,
at $z=0.3, 0.4, 0.5$. 
This last condition is imposed to remove extremely marginal
peaks from consideration.  
When a candidate meeting the above criteria is found, we determine
its estimated redshift by finding the maximum in the \lfine\ versus
$z$ curve at the maximum likelihood sky position.

Once candidates meeting the above criteria
are found, we subject them to two further tests. First,
the image(s) are inspected at the position of the
cluster candidate.
An overdensity on a spatial scale appropriate to
the estimated redshift must be visually apparent.
Second, we produce a background-subtracted luminosity histogram for the
putative cluster within $r_{core}$, $2r_{core}$, and $4r_{core}$.  
A reasonable excess approximating a Schecter function 
must be apparent to the eye and in the local luminosity function 
if the redshift and richness estimates are to be believed.
Although such visual tests subjective,
they are still necessary if artifacts are to be avoided.
As will be demonstrated in \S 4,  
most candidates identified with the automatic criteria
pass the subjective visual tests.  

\section{A Worked Example: First \stacs\ candidates and confirmed clusters}
\newcommand{\ra}{{\rm RA\/}}
\newcommand{\dec}{{\rm DEC\/}}
\newcommand{\hr}{^{h}}

\begin{figure*}
\begin{minipage}{180mm}
\includegraphics[scale=0.95, angle=0]{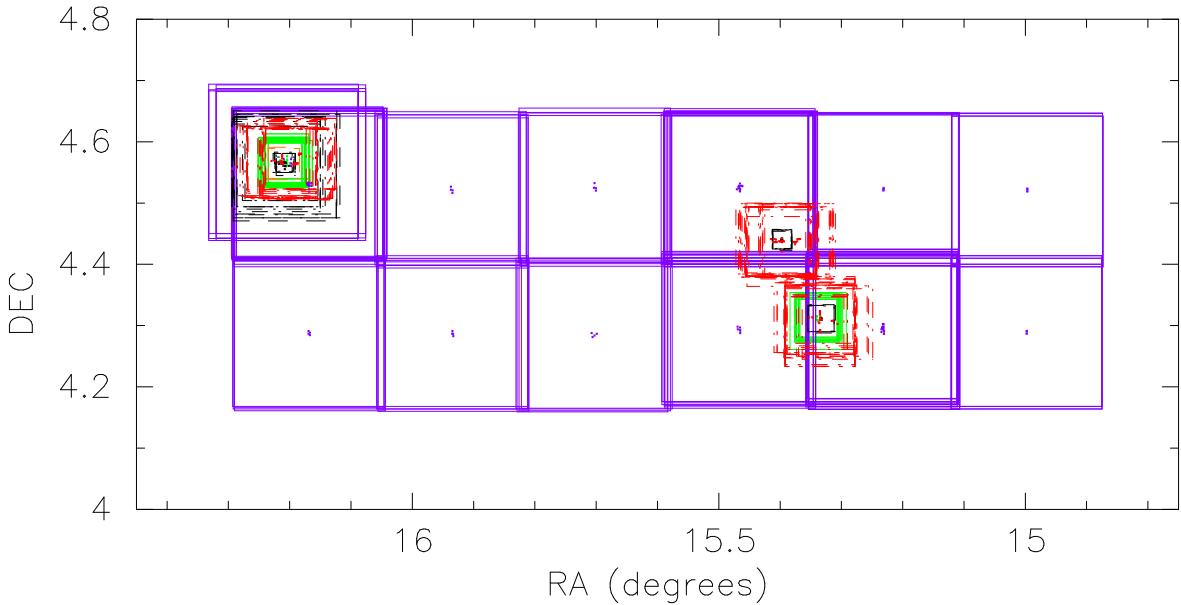}
\caption{{\small Positions of the CCD images obtained
by the SCP in the 01+04 field (see text).  
The regular
array of blue squares (CTIO prime focus frames taken in October
and November 1995) have been processed with the 
full \stacs\ analysis procedure; the others were supernova follow-up
frames and are not useful for \stacs .}}
\label{fig:f0100}
\end{minipage}
\end{figure*}

Our first step, as noted in \S~3, is to search the SCP data base
for sets of images that coninuously cover a reasonably large
patch of sky---a few tenths of a square degree or more. An
example of such a set of images is shown in
Figure~\ref{fig:f0100},
which shows all SCP frames covering a patch of sky centered on
$\ra \approx 1\fh 0,$ $\dec \approx 4\fdg 4$ (hereafter called 
the 01+04 field).  
The large blue boxes represent CTIO prime focus frames, which
are the ones used in our analysis. The smaller boxes represent
follow-up images, acquired at other telescopes, with smaller
fields of view, that we do not use; the small gain
in image depth achieved does not warrant
the significant additional effort
required to co-add images from different telescopes. 

\subsection{Production of the Galaxy Catalog}

At each of the 12 frame positions shown in
Figure~\ref{fig:f0100},
we combine the individual CCD images
to produce a single deep frame with
a typical effective exposure time of $720\, s.$ 
For each of the 12 deep frames thus generated, 
we determine the best sky value, run FOCAS
(tasks {\sc setcat, detect, sky, skycorrect, evaluate}, and
{\sc splits})
with the sky and sky noise fixed, 
using a detection threshold of $2\times \sigma_{sky}$ and a
minimum detection area of 4 pixels, 
use IRAF tools to identify 10-20 stars to 
set the FOCAS catalog point spread function 
with {\sc setpsf}, and finally run the FOCAS task {\sc resolve}.
This results in a catalog of galaxy and star 
positions and magnitudes for each frame. 

The photometric zero point
for the magnitudes is obtained from the SCP data base and is
accurate to $\sim 0.1$ mag or better. 
With the established FOCAS identifications we repeat the 
object photometry with a more sophisticated algorithm
using local software.  
The individual catalogs are
then mapped to celestial coordinates by matching objects
within them to the USNO astrometric catalog (we do not
discriminate between stars and galaxies for this
procedure, as many USNO catalog soures are galaxies).
The individual catalogs, now photometrically and 
astrometrically calibrated, are then
combined to form a single catalog for the entire imaged
region shown in
Figure~\ref{fig:f0100}.
Although there
is significant spatial overlap among the deep frames,
our combination algorithm ensures that very few duplicate
objects remain in the final catalog and that the 
exclusion boxes are propagated to the final catalog correctly.
One final step is to correct for galactic extinction 
using the \citet{schlegel98} dust emission maps and extinction
calibration\footnote{Software and data were obtained from
the ftp site deep.berkeley.edu:/pub/dust/maps}.  Although most
of the SCP fields were located in regions of low extinction,
correcting for it changes the estimated $z$ and $\calr$
values and in some cases noticeably corrects galaxy density variations
caused by dust.  

\begin{figure*}
\begin{minipage}{180mm}
\includegraphics[scale=0.95]{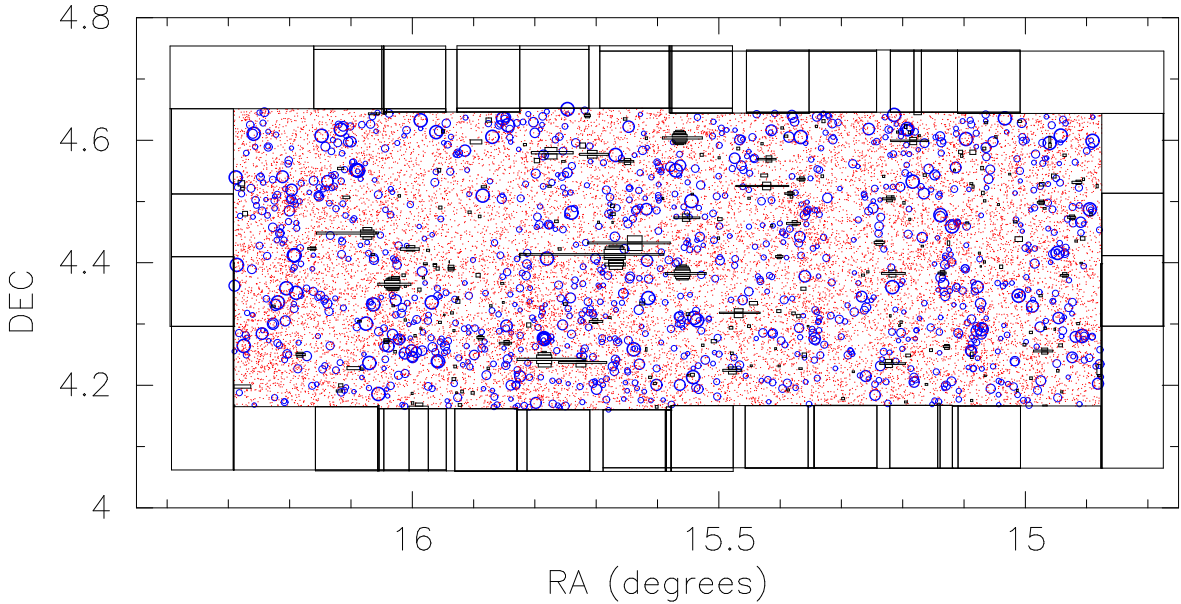}
\caption{{\small Positions of a subset of the galaxies
detected in the CTIO prime CCD frames shown in
Figure~\ref{fig:f0100}.
Blue markers are sized according to the magnitudes of those galaxies
with $m_R\leq20$, and galaxies with $20 < m_R < 23$ are shown with red
dots.  Exclusion boxes are in black.}}
\label{fig:galpos}
\end{minipage}
\end{figure*}

Figure~\ref{fig:galpos}
shows the positions of bright
($m_R \le 20,$ approximately 1200 objects) 
and faint ($20.0 \le m_R \le 23.0,$ approximately 22,000 objects) 
subsets of the galaxies in the final catalog.
(For clarity, an additional $\sim 15,000$ galaxies with $m_R > 23$ 
are not shown in the Figure.) 
The bright subset is shown by blue circles with size proportional
to brightness; the brightest galaxies have
$m_R \sim 17.$ The faint subset is shown as (red) dots of fixed size.
The exclusion boxes are in black.  The boxes in amongst the galaxies 
generally indicate the effects of saturated stars, while
the boxes around the periphery 
identify for \clusterfind\ the edges of the galaxy catalog.  

\subsection{Determining the Background Galaxy Distribution}

The next step is to determine the parameters of the power-law
distribution of background galaxies, Eq.~(\ref{eq:fielddist}) above.
Note first that if we adopt $\theta_f = 1\farcm 0$ and take
$l_f$ to be the flux corresponding to $m=20,$ Eq.~(\ref{eq:fielddist})
is equivalent to the statement that the total number of catalog
galaxies brighter than apparent magnitude $m$ is 
\be
N(<m) = \frac{N_f\,\Omega_{{\rm eff}}}{\beta-1}\,10^{0.4[\beta-1](m-20)} \,,
\label{eq:nltm}
\ee
where $\Omega_{{\rm eff}}$ is the effective solid angle of the catalog
(i.e., the total area minus the combined areas of patches where
bright stars precluded galaxy detection). Thus, the slope
of the $\log N(<m)$ versus $m$ graph is $0.4(\beta-1),$ while
the amplitude of the graph determines $N_f.$ In particular,
for $\beta\approx 2$ (see below), 
$N_f$ is essentially the number of galaxies per square
arcminute brighter than $m=20.$ 

In practice the power-law distribution of field galaxies
is not realized over the entire range of apparent magnitudes. 
At the bright end, the number counts
are biassed by small number statistics
and the loss from the catalog of galaxies
brighter than $m_R\sim 16$ due to detector saturation, 
and can either exceed or fall short
of the power-law expectation. At the faint end, incompleteness
sets in above $m_R \sim 23.$ 
To quantify this, we fit the power-law
only over a range of magnitudes, $19 \simlt m_R \simlt 22.5,$
with the exact range chosen on a case-by-case basis.
Brighter than this range, we do not fit the
number counts. Fainter than the chosen range, we multiply
the power law by an incompletness function, which we take
to have a Fermi-Dirac form,
\begin{equation}
g(m;m_c,\Delta m) = \left[ 1 + e^{(m - m_c)/\Delta m} \right]^{-1} \;,
\label{eq:incompl}
\end{equation}
where $m_c$ is a characteristic cut-off magnitude and $\Delta m$ 
represents the sharpness of the cut-off. 
Note that we use the exclusion boxes at this stage both to remove
from consideration spurious objects and to measure properly
the area of sky observed.  Processing the catalog without 
the exclusion boxes
gives inconsistent results, both with respect to the number density
normalization and the shape of the distribution at the faint end.  

\begin{figure}[t!]
\centering
\includegraphics[scale=0.45]{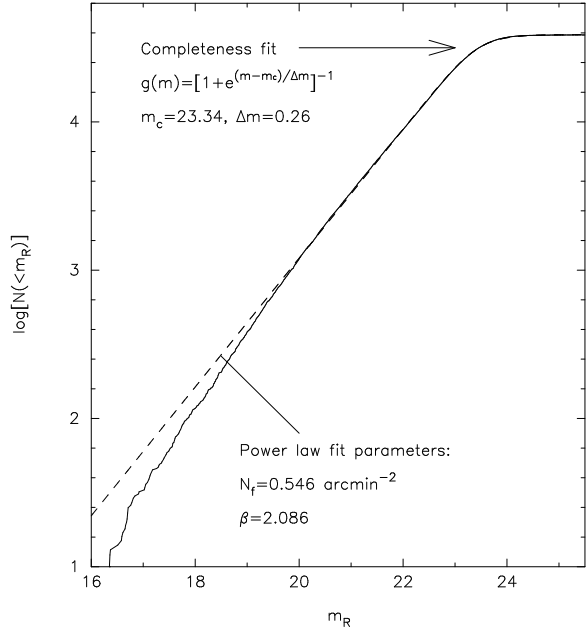}
\caption{{\small Total number of catalog galaxies as a function
of apparent magnitude, for the SCP fields shown in the
previous figures. The solid  line represents the
measured counts, while the dashed line is the fitted
power law (Eq.~\ref{eq:nltm}) multiplied at the
faint end by the incompleteness function (Eq.~\ref{eq:incompl}).}}
\label{fig:back1}
\end{figure}

Figure~\ref{fig:back1}
shows the results of carrying out
this procedure for the catalog derived from the frames
shown in
Figure~\ref{fig:f0100}.
The fitted parameters for
the power-law distribution, $N_f$ and $\beta,$ are
indicated on the figure, as are those 
of the incompletness function. 
The indicated values of $N_f$
and $\beta$ are consistent with what we find in other SCP fields,
though they vary from field to field, possibly due to 
errors in the photometric zeropoint or
variations
in seeing causing slight differences in the efficiency of star-galaxy
identification.  
Note that
the width of the incompleteness cutoff, $\Delta m \approx 0.25,$
is quite small. For this reason, when we apply \clusterfind,
we consider only galaxies with $m \le m_c,$ and assume
that the formalism of a strict magnitude limit, implicit
in Eqs. (\ref{eq:lfine}--\ref{eq:solveforlambda}), 
is a good description of the data.
This simplification means \clusterfind\ loses some signal
we could potentially gain by including galaxies just
past the faint limit, and it also means the equation
for the background galaxy distribution overestimates
the catalog number density just short of the faint limit.
An alternative would be to use all the galaxies, and to 
incorporate the incompleteness function, Eq.~(\ref{eq:incompl}),
into the MFA formalism. However, this would complicate the
mathematical expressions involved as well as introduce  a higher
fraction of 
spurious objects, and we have judged it
not to be worth the added complexity. 

\subsection{Identification of cluster candidates}

The next step is to apply \clusterfind\ to the
galaxy catalog. This requires that we adopt values for the
model parameters on which the algorithm of \S 3
depends; our choices are shown in
Table~\ref{tab:clust_params}.  Only the field galaxy distribution
parameters $\beta$ and $N_f,$ and the magnitude
limit $m_{lim},$ are newly determined
for each catalog. The other parameters, in particular those
describing the cluster properties, are hardwired to
the values given in Table~\ref{tab:clust_params}. The cluster galaxy
luminosity function
was estimated from data in \citet{driver98} based on our assumed
cosmology.  We execute the algorithm using observed quantities;
the final result does not depend upon $h$ except through the
explicit \mstar, $r_{core}$, and $r_{max}$ dependencies listed in
the table.  
We have adopted the $r_{max}$ used by  \citet{post96} and
\citet{kep99}, but have used a larger $r_{core}$.

The K-correction is problematic for the SCP data, since the fields
primarily consist of $R$-band images and we cannot estimate a spectral
type except for low redshift galaxies in the sample.  We therefore
have applied a mean K-correction to all the galaxies equally based 
on an average of functions taken from \citet{cole80}.  While this
will give increasing errors in galaxy magnitudes with redshift, we
note that the detection of a cluster is based on the combination of
the local apparent luminosity function excess  and 
the angular scale size of the overdensity, but the latter is
weakly dependent on $z$ at high redshifts (and moreover depends on both
the specific cosmology and possible evolution in the scale size of
clusters, both of which we have fixed).  
Therefore, a systematic error in the adopted K-correction 
will give systematic but small errors in predicted $z$.  Because we
do not intend to do science with the un-confirmed cluster 
catalog, the estimated $z$ values will be used only to prioritize
the targets and to chose the best spectroscopic configuration
(choice of grism, filter, etc.).  

\begin{figure*}
\begin{minipage}{160mm}
\includegraphics*[scale=0.75]{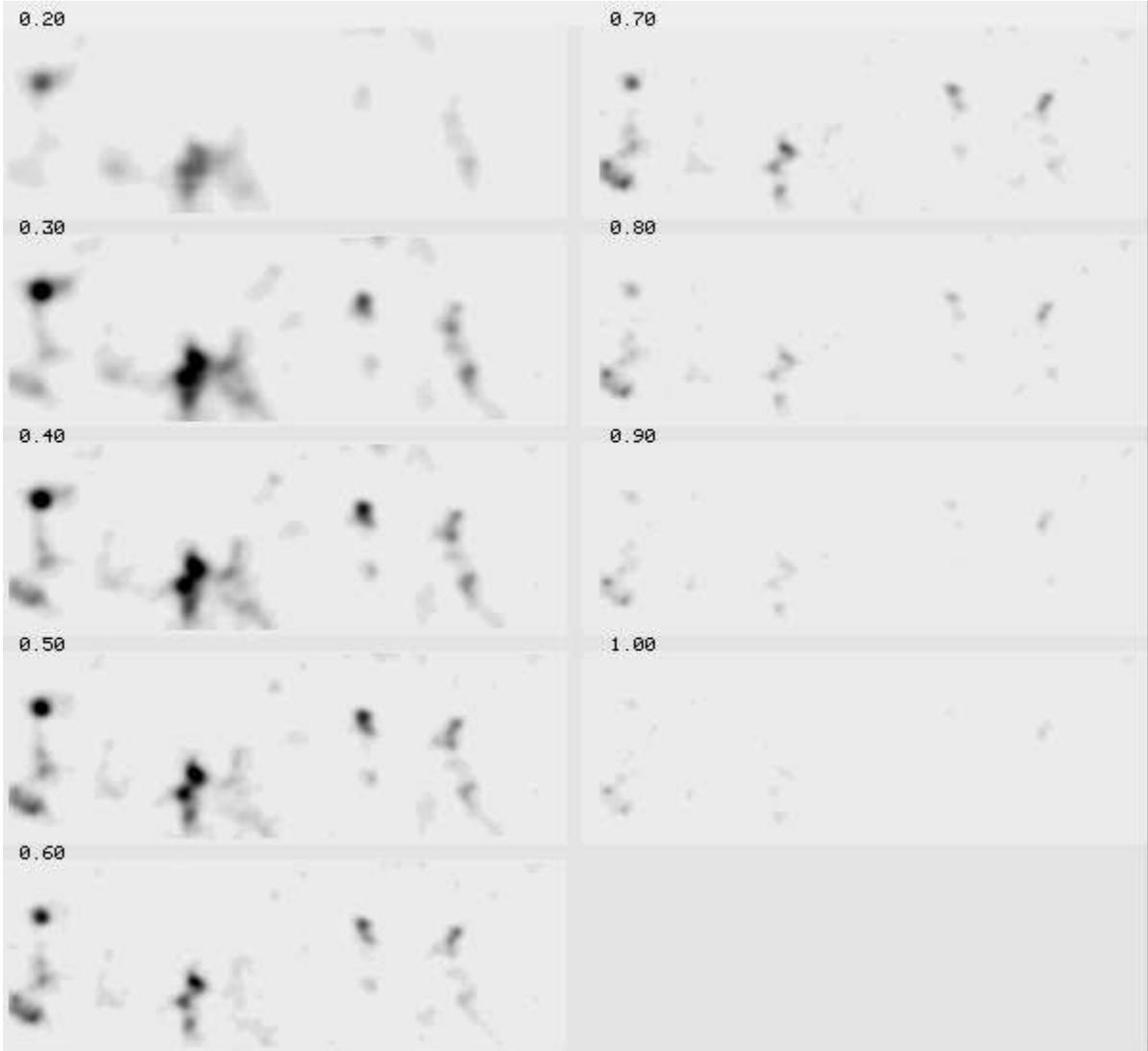}
\caption{{\small The 3-dimensional $\lfine$ map resulting from 
a run of \clusterfind\ on the 01+04 field 
galaxy catalog.  The redshift for
each RA-DEC plane is shown in the upper left hand corners.  Values
of $\lfine$ are suppressed (set to -1 in the plot) where $\calr < 0$.  
The confirmed cluster J0104.8+0430 is represented 
by the strong, nearly circular peak in
the upper left hand corner.}}
\label{fig:lfine0100+04}
\end{minipage}
\end{figure*}

\begin{figure}[t!]
\centering
\includegraphics[scale=0.40]{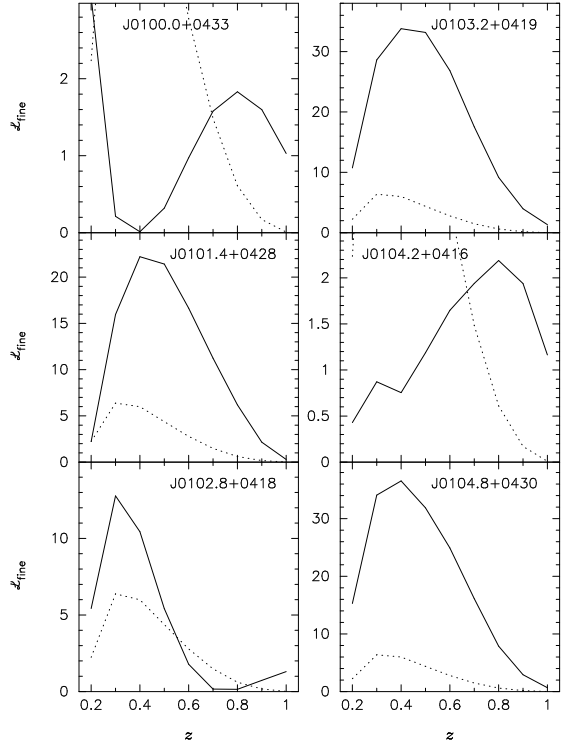}
\caption{{\small Likelihood $\lfine$ versus redshift
for six candidate clusters obtained from the
01+04  
field. The peak of
each curve indicates the most probable redshift
of the candidate cluster, and the richness at
the peak its estimated richness. The dotted lines
represent the 95$^{\mbox{th}}$ percentile \lfine\ 
value as a function of redshift, used
as a threshold (see text).}} 
\label{fig:lfinevsz}
\end{figure}

\begin{figure}[t!]
\centering
\includegraphics[scale=0.44, angle=0]{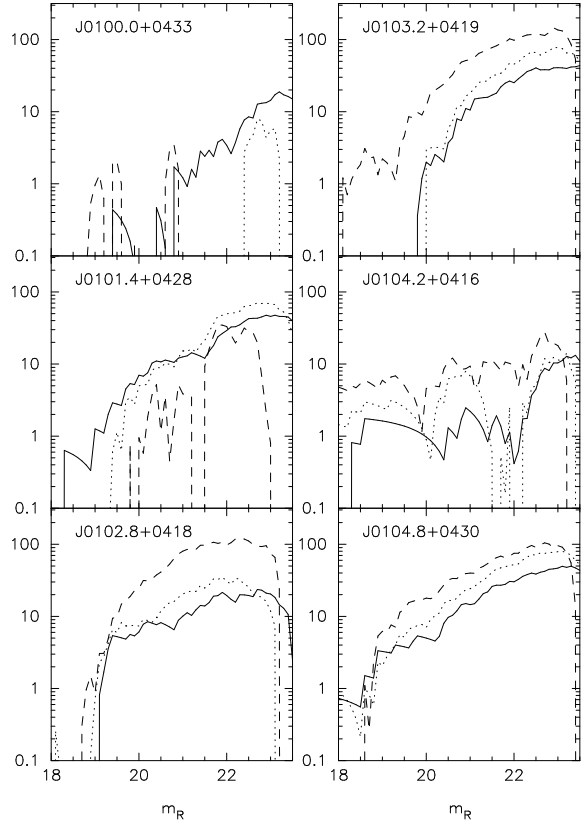}
\caption{{\small Luminosity functions for the candidates
indicated in
Figure~\ref{fig:lfinevsz}.  
The horizontal axes are $R$-magnitude and the
vertical axes are the integrated excess number counts
($N(<m) - N_{background}(<m)$) within circles centered on
the cluster candidate.  
The solid lines indicate the counts within $r_{core}$ of
the center, dotted within $2r_{core}$, and dashed within $4r_{core}$.  
The background distribution is assumed to be the straight power
law used in \clusterfind , and  exclusion boxes were ignored.}}
\label{fig:lumfunc}
\end{figure}

As noted in \S 3, the output of \clusterfind\ consists of maps
of the likelihood $\lfine$ and richness $\calr$ on the plane
of the sky ($\alpha,\delta$) and at each of 9 redshifts,
$0.2,...,1.0.$ 
Figure~\ref{fig:lfine0100+04}
shows the $\lfine$ map for the region in question.  
The likelihood map is searched for peaks
that lie above a 95\% threshold (i.e., that have $\lfine$
in the top 5\% of values {\em at that redshift}) at three
consecutive redshift planes.  In
Figure~\ref{fig:lfinevsz}
we plot, for six such peaks derived from the 01+04 
field, $\lfine$ versus redshift.  In
Figure~\ref{fig:lumfunc}
we plot the local excess apparent luminosity function for the
same peaks within concentric circles around the candidate position.  

Table~\ref{tab:clustercat} shows the candidates from the
01+04    
field that pass the
tests noted.  

\subsection{Matched filter algorithm limitations and tests}

Of 20 cluster candidates
found with the MFA listed in  Table~\ref{tab:clustercat}, 
11 appear to be associated with larger scale 
galaxy overdensities or substructure, largely confirmed by
their apparent luminosity functions (see
Figure~\ref{fig:lumfunc}).   
There are several reasons to be careful in the
interpretation of MFA results without spectroscopic confirmation.
First, the non-uniform background galaxy distribution 
effectively changes the 
threshold level \citep[see][]{bramel00}, and since
the \lfine\ map is nonlinear, the rise of a local maximum above the
surrounding \lfine\ value is not necessarily indicative of what the
cluster's signal would be if seen against the mean background
density. This may be somewhat ameliorated if we were to use a locally
determined threshold when analyzing the \lfine\ map, but 
nonlinearity in the \lfine\ calculation would still
give incorrect values for detected cluster candidates. 
Another approach is to use a local background galaxy
distribution when calculating \lfine, but estimation of the
background counts will be problematic, because of similarity
in the scales of background density variations and cluster
angular sizes.  

A second  problem with the MFA is that it assumes real clusters 
have a strongly
peaked spatial distribution and there are no significant overdensities 
that are not clusters.  Overdensities in the form of ``sheets'' and ``strings''
will also contribute to the \lfine\ signal, and with projection
effects we may identify candidates that are not massive bound structures.  
A third problem when searching for medium to poor clusters is that 
overlapping cluster candidates are common.  The model for \lfine\ is
a single cluster on a uniform background, and the nonlinearity in the
\lfine\ value prevents effective deconvolution. This may, however, be a
minor concern since cluster characteristics vary enough that such
deconvolution could not be used even with a linear likelihood
\citep[such as \lcoarse\ from][]{kep99}.
In such cases a thorough spectroscopic redshift campaign is required;
such \lfine\ signals may indicate merging clusters or the
projection of large scale structure elements.

We have taken the position that the MFA, particularly when used
with single-band images, will not have sufficient information to identify 
true clusters with small and well-understood false positive rates and 
completeness properties.  We therefore intend to use the MFA only to
identify candidates, emphasize completeness over avoiding false positives,
and confirm the clusters spectroscopically. It is further
evident that we cannot reliably identify true clusters in some cases without
thorough investigation of the velocity field to reject cases where
projection of large scale structure, rather than gravitationally
condensed clusters, provides the galaxy overdensity detected by \lfine.
We also expect to find high redshift clusters projected behind low
redshift clusters, and these may be unidentified without thorough 
analysis of the velocity field to faint limits. 
Merging clusters or collapsing proto-clusters may also show a large 
velocity dispersion, given a sparse sampling of galaxy redshifts,
and give us an incomplete picture of the candidate.  
The use of deep images to find candidates is only the first step.  

Since the matched filter algorithm implements a specific model
for a cluster's spatial galaxy distribution and luminosity 
function, it is reasonable to ask whether our detection efficiency
is model-dependent.  We have performed a preliminary analysis of
this question by investigating the filter's response to simulated
clusters over a wide range of the parameters listed in Table 1.
The ratio between a cluster's peak $\lfine$ value and the 95th
percentile $\lfine$ value (our detection threshold), appears robust
under reasonable variations in the assumed background and member
galaxy distributions. The real variations in cluster properties
are therefore not likely to be an obstacle in assessing this
survey's completeness. Differences between the filter model and
the properties of our simulated clusters do create systematic
errors in the algorithm's richness and redshift estimates, but
these values are only used to prioritize our candidates for
follow-up observations.

One fortuitous check on our algorithm is provided by
MS~1054.4-0321 \citep{gioia90,stocke91},
re-discovered as STACS~J1056.9-0337.  At the time of
those particular SCP observations (prior to March 1997), the 
choice of fields sometimes included known high redshift clusters to
boost the probability of finding supernovae, so this re-detection
cannot be included in a formal \stacs\ statistical sample.  It can, however, 
be used to check our parameter estimation accuracy.  Our algorithm
found $z_{est} = 0.7$, $\calr = 83.5$, and the peak \lfine\ value was
a factor of 16 above the (95th percentile) threshold.  
\citet{tran99} found $\langle z\rangle = 0.833$, indicating our estimated
redshift is more than one $z$-step low.  Our \calr\ represents
a direct fit to an observable, and therefore is a valid measure
of the cluster, but the number quoted is based on $z_{est}=0.7$;
going  back to the \calr\ map and interpolating we
find $\calr (z=0.833) \sim 160$.  Use of \calr\ is a good simple method
of estimating the optical luminosity, since it subtracts the background
galaxy distribution and it weights the galaxies
in both radius and magnitude according to our Schechter function + Plummer
law cluster model rather than having hard cutoffs in magntiude and
radius.  However, real clusters are not round (this one manifestly so), they
may be projected against an over- or underdense region of background
galaxies, and they may not have a Schechter luminosity function (e.g.
may have a CD galaxy at the center which adds greatly to the luminosity 
but does not affect \calr\ proportionally), so a proper luminosity
measurement must identify cluster members more accurately.  

The underestimate of $z$ for high redshift clusters may be systematic in
origin. This might come from a K-correction error (an overestimate of
the effective value), an error in the assumed cosmology (this
could account more than half a magnitude, cf. Figure 1
of \citealt{perl99},
which can give errors of $\sim 0.1$ in redshift),
or an evolving luminosity function.  
As the catalog of confirmed clusters grows we will be able to
determine whether such errors are systematic or random.

    As expected, MS1054.4-0321 is the richest cluster dectected 
out of  $\sim 3\,\sqdegs$ processed.  The encouraging aspect is
that much poorer clusters have significant \lfine\ signal as well, 
even with our single filter images.  We have tens of candidates with
optical luminosities (as indicated by $\calr$) within a factor of
a few of MS1054.4-0321, and we can expect deep optical surveys such
as \stacs\ to  have a high degree of completeness at mid- to high
richness.

\subsection{Spectroscopic results}

\begin{figure}
\centering
\includegraphics[scale=0.45]{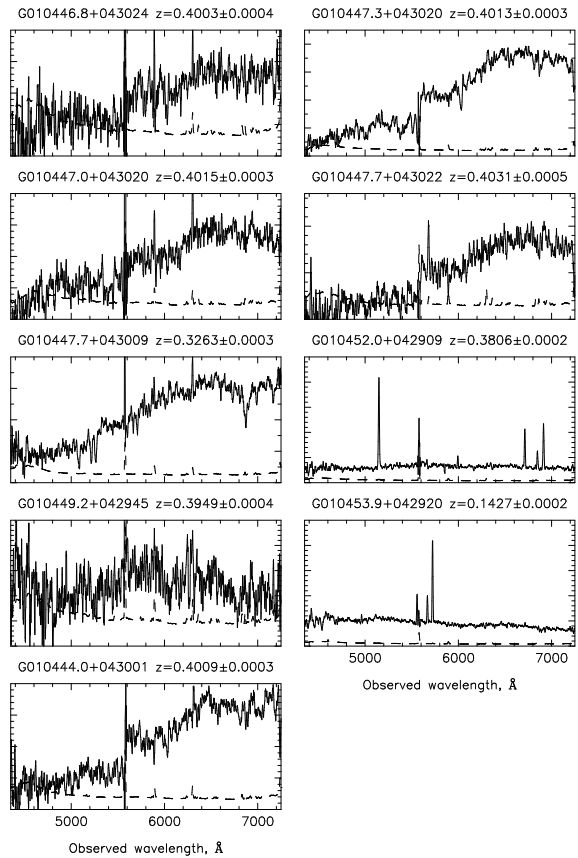}   
\caption{{\small Spectra of galaxies in the field of STACS J0104.8+0430.
The solid lines are the spectra, smoothed with a 5-pixel boxcar 
filter (approximately the size of the resolution element), and 
the dashed lines are the approximate pixel uncertainties.  
The verical axes are roughly calibrated F$_\lambda$. The H+K
absorption features are just short of the $\lambda 5577$ sky
emission line subtraction residuals in the cluster
members.   }}
\label{fig:galspecs}
\end{figure}

    During the first season of science operations of the HET, we
confirmed a cluster candidate in the 01+04 field 
with the Marcario Low Resolution Spectrograph \citep{hill98}
in longslit mode.
The spectra were taken in October 1999 with 2 arcsecond slits 
(resolution $\sim 600$)
at various position angles allowing several galaxies per exposure.  
Figure~\ref{fig:galspecs}
shows the spectra, including 5 cluster members,
1 likely member, and 3 unassociated galaxies.  The data were
reduced with IRAF {\sc apextract} tasks and the redshifts measured
with the {\sc rvsao} package using SAO galaxy spectrum
templates, although we have found 
that the galaxy templates from \citet{kinney96} are better 
for higher redshift galaxies due to their broader wavelength coverage.  
The mean redshift $\langle z\rangle=0.400$ 
of the 6 likely members is
close to the $z=0.4$ value predicted by \clusterfind.  

    The current list of confirmed clusters is in
Table~\ref{tab:clusterconf}. We have adopted a liberal criterion
for the confirmed candidate list, requiring only 3 redshifts within
2000~\kms. Columns 6 and 7 give the number of redshifts in that range
and the total number of redshifts that we know in the field,
respectively; these numbers are indicative of the strength of our
spectroscopic confirmation. We recognize that some of
these candidates may indeed be rich groups, projections of 
large scale structure, or mere coincidences; we may also find 
a foreground group's redshift has been determined instead of
a true background cluster. Our target confirmation level for
HET spectroscopy is to obtain on the order of a dozen member
redshifts for each candidate before adding it to the \stacs
cluster catalogue. Extensive spectroscopy is required to confirm 
a virialized, massive structure, especially for poor and high
redshift clusters where the background contrast is low. Follow-up
observations of the confirmed catalogue are essential for most
of our science goals.

    An online version of the \stacs\ confirmed cluster catalog
can be found at \linebreak[4]
{\tt http://redshift.stanford.edu} and will be updated as we
obtain further HET data.  We will also make available upon request
the positions, redshifts, and spectra of galaxies in these fields.
Those who would like to study these clusters further should 
contact KLT.  As indicated in the introduction, \stacs\ may turn
out to be more limited in scope than our original plan, and
we will therefore not guard our data closely. The candidate list
and software will be made available to any group wishing to continue
the search or make use of our catalogue, which contains a wide
variety of interesting high-redshift cluster candidates.

\section{Conclusions}

The purpose of this paper has been to describe \stacs, a new
research program aimed at identifying previously unknown
intermediate and high redshift galaxy clusters from moderately
deep, archival CCD images covering several tens of square degrees
of sky. These images were originally obtained by the Supernova
Cosmology Project \citep{perl99} for the purpose of finding
high-redshift supernovae. Our short term goals are to identify
new clusters with a wide variety of masses in these images by 
using a matched filter algorithm that provides reasonably accurate
estimates of cluster richnesses and redshifts. We then follow up
viable candiates with HET spectroscopy to measure redshifts and
confirm that the objects are bona-fide clusters. Our long-term goal
is to use the clusters as cosmological probes, and in particular to
measure the cosmological density parameter $\omegam$ and the amplitude of
density fluctations to $\sim 10\%$ accuracy. This long-term goal cannot
be carried out using the archival images and HET data alone,
because it requires accurate estimates of the clusters' virial
mass, for reasons outlined in \S 2.  Additional observations
will thus be needed, including multiobject spectroscopy at the
HET and other 8 m class telescopes. If possible, we will also
pursue X-ray, weak lensing, and Sunyaev-Zel'dovich observations, 
either ourselves or through collaboration, to measure cluster
mass profiles. These observations will require the efforts of
many workers, and we thus plan to make our cluster candidates
available on the Internet for follow-up by the scientific community.

In this final section, we further discuss some recent key 
issues that will affect \stacs\ and similar projects.

\subsection{Press-Schechter and Beyond}

As noted in \S 2.4, the familiar Press-Schechter formalism
for predicting the cluster abundance $n(M,z)$ has now
been shown clearly to fail at the low and high mass ends
\citep{g99,j00}, and more accurate analytic expressions
for $n(M,z)$ have been proposed and tested \citep{st99,smt99}.
Fitting formulas to the mass functions found in large-scale
simulations are also becoming available \citep{j00} as an
alternative to traditional models.
There has not yet been a detailed study of the effect of
changes in the model cluster mass function on cosmological
constraints. The most comprehensive assessments of the power
of the cluster abundance tests done to date are probably those
of \citet{rvlm99} and \citet{hmh00}, both of which assumed the
validity of PS in arriving at their findings. \citet{hmh00},
in particular, argued that extremely accurate constraints
($\sim 1\%$) on $\omegam$ could be obtained from a $400$
square degree X-ray survey of clusters, even when variations
on the nature of the cosmological constant (``quintessence'')
are allowed. This high accuracy stems, in part, from the assumption
that the theoretical predictions hold well into high-mass,
high-redshift ($z > 1$) regimes where PS apparently breaks down.
It is not yet know whether the improved analytic formulae of
\citet{st99} and \citet{smt99} will fully withstand scrutiny, in
this regime of rare objects, as new N-body results come in.

We suggested in \S 2 that such uncertainty can be at least
partially alleviated by focussing our efforts on the mass
and redshift ranges where both the PS and the newer predictions
are accurate. This means directing the abundance
test at objects for which $\delta_c(z)/\sigma_M$ is not very
large---which in practice means moderate cluster masses
($M\simlt 5\times 10^{14}\,\msun$) and redshifts
($0.3 \simlt z \simlt 1.0$). Such objects are, in any case,
far more common than the high-mass, high-redshift
clusters which could prove definitive if the PS
formulae were rigorously correct. They are also diffulcult
to detect in quantity through any means but an optical survey.
Still, it is critical for the success of the cluster abundance
test of cosmology that a better theoretical understanding
of the evolution of $n(M,z)$ be reached. 

\subsection{Ancillary Scientific Goals}

While surveys such as \stacs\ are aimed mainly
at constraining cosmological parameters, especially
$\omegam,$ this is not and should not be their only objective. 
It may turn out that other approaches to cosmological parameter
estimation may prove more fruitful, although cluster 
evolution still provides a valuable independent constraint.
The recent release of CMB anisotropy power spectra
by the {\sc Boomerang\/} \citep{L00} and {\sc Maxima\/} 
\citep{b00} teams has demonstrated the potential
of CMB observations to carry out precision cosmology,
and the coming year will see an order of magnitude
improvement in these results with the Microwave Anisotropy
Prove (MAP) satellite. If indeed CMB experiments
live up to their potential and convincingly measure  
the cosmological parameters\footnote{And one needs to bear
in mind that this is not a foregone conclusion; surprises
may yet await us, as underscored by the absence of a detectable
secondary peak in the {\sc Boomerang\/} data.}, what new insights
will cluster surveys provide?

With $\omegam$ and $\sigmaeight$ already determined, a measurement
of $n(M,z)$ out to $z \approx 1$ will be crucial in testing
our theories of structure formation. Will $n(M,z)$ turn out
to match N-body simulations, which will now have no remaining
cosmological degrees of freedom? If not, the cluster abundance
may be telling us something about a breakdown of other
key assumptions about structure formation, such as
Gaussianity of the initial conditions \citep[e.g.][]{w00}
or the dominant role of gravity in large-scale structure formation.

In addition, rich clusters of galaxies are excellent ``cosmic
laboraties'' in which to test ideas about dark matter. 
Their mass distributions can be studied using X-ray, Sunyaev-Zeldovich
(SZ), weak lensing, and redshift observations, and may be
compared with the expected forms of dark matter halos.
The fact that the halos around small galaxies do
not appear to match the expectations from CDM simulations
has been cited as evidence for self-interacting dark 
matter \citep{sperg00} and for warm rather than cold dark matter 
\citep{hogan00}. However, surprisingly little is known about
the dark matter distribution in clusters, and if such ideas
are to be tested, we will need to learn much more about cluster
halos. Cluster surveys such as \stacs\ will identify many
new objects to study, at a wide range of masses and redshifts,
that will bear on this important problem.

Indeed, as discussed in \S 2.3, a better understanding of
cluster mass profiles is critical not only as an ancillary
goal of \stacs, but to its primary goal of comsological
parameter estimation as well. Ultimately one must estimate
the virial masses of clusters in order to use them as
cosmological probes, because it is in terms of
virial mass that N-body experiments
and analytical formulae, Press-Schechter of other, ultimately
make their predictions. Because virial masses represent
the mass enclosed within a given overdensity, some
knowledge of the density profile is needed in order
to go from observational data to virial mass. 
It has often been assumed that a single temperature
measurement from X-ray data enables a direct
inference of virial mass using simple, redshift-dependent
formulae. This may be true in the future, but much
work remains to be understood about the origin and
evolution of the intracluster medium before such 
a mass-temperature relation can be applied with
confidence \citep{math01}. Similarly, weak lensing and
velocity measurements of a cluster require a mass model to
be converted into virial mass \citep{w00}. A central goal of
cluster surveys should be follow-up observations, using
a variety of techniques, to constrain mass profiles
and thereby obtain unbiased virial mass estimates.
This has not been done in cluster surveys to
date, and this may be one of the reasons for the
discrepant cosmological conclusions drawn by the authors
cited in \S 1 above. 



\section*{}

We gratefully acknowledge the assistance of many: 
at LBL R. Quimby, G. Aldering, numerous others; 
at Stanford S. Church, R. Romani, V. Petrosian for useful discussions, 
students D. Sowards-Emmerd,
D. Grin, F. Tam, E. Young, Y. Dale who contributed to software 
and/or data processing; 
at Univ. Sternwarte G\"{o}ttingen: F. Hessman; at NOAO F. Valdez, 
M. Fitzpatrick;
at McDonald Observatory special thanks to 
resident astronomers Grant. Hill, M. Shetrone.

The LRS is named for Mike Marcario of High Lonesome Optics who fabricated
several optics for the instrument but died before its completion.


\setcounter{figure}{0}
\clearpage

\clearpage

\begin{deluxetable}{lll}   
\footnotesize
\tablecaption{Clusterfind Parameters. \label{tab:clust_params}}
\tablewidth{400pt}
\tablehead{
  \colhead{Parameter} & \colhead{Value or range} & \colhead{Description}}
\startdata
$\beta$   & $\sim 2.0$   & bckgd.\ dist. power law index\tablenotemark{a}   \\
\nfield   & $\sim 0.6$  & bckgd.\ density (arcmin$^{-2}$) at 
    $m_R=20$\tablenotemark{a} \\ 
\mlim     &$\sim 23.2$--$23.6$&limiting magnitude 
    ($\Leftrightarrow l_{min}$)\tablenotemark{a} \\
\mstar    & $-21.55 + 5\log{h}$ & Schechter fcn.\ \lstar\ absolute R 
    magnitude\tablenotemark{b,c}  \\
$\alpha$  &  1.125       & faint end slope of Schechter 
    fcn.\tablenotemark{b} \\
\rcore    &  0.25$h^{-1}$ Mpc  & core radius of cluster\tablenotemark{c,d}  \\
\rmax     &  1.00$h^{-1}$ Mpc  & truncation radius of cluster\tablenotemark{c,d}  \\
$(\omegam,\omegal)$&$(0.3,0.7)$&cosmological parameters 
    (for calculating $d_A,$ $d_L$)  \\
gridspace &  0.3 arcmin    & spacing of grid points  \\
E/S0:Sab:Sbc:Scd & 4:3:2:1 & ratio of galaxy type contribution to 
                              K-correction\tablenotemark{e} \\ 
\enddata
\tablenotetext{a}{The value determined separately for each galaxy catalog.}
\tablenotetext{b}{Schechter fcn. for clusters: 
   $N(l) \propto (\frac{l}{l^\ast})^{-\alpha}e^{-l/l^\ast}$, where
   $l^\ast = L^\ast/4\pi d^2_L + {\rm ext.\ corr.} + {\rm K-corr},$ 
   where $d_L$ is luminosity distance.}
\tablenotetext{c}{$h\equiv H_0/100\kmsmpc$}
\tablenotetext{d}{Modified Plummer law: galaxy surface number density
   $F(\theta) \propto [1 + (\theta/\theta_{core})^2]^{-1} 
   - [1 + (\theta_{max}/\theta_{core})^2]^{-1}$,
   where $\theta$ is the angle on the sky to
   the cluster center, and $\theta_{core}$ and $\theta_{max}$ are the
   angles corresponding to \rcore\ and \rmax\ cluster 
   radii. }  
\tablenotetext{e}{The mean K-correction $k = \sum w_i k_i(z)$ where 
   $k_i(z)$ are the K corrections for each galaxy spectral type 
   as a function of $z$ from \citealt{cole80} and
   $w_i$ are the corresponding weights listed in the table (normalized).
   see text.}
\end{deluxetable}

\clearpage


\begin{deluxetable}{lrrrrl}   
\footnotesize
\tablecaption{Cluster candidate list for the 01+04 field.
    \label{tab:clustercat}}
\tablewidth{430pt}
\tablehead{
\colhead{Name\tablenotemark{a}} & 
         \multicolumn{2}{c}{Coordinates, J2000} & 
         \multicolumn{2}{c}{Estimated parameters} & 
         \multicolumn{1}{c}{Notes:} \\
\cline{4-5} \\
 & RA (deg.)  & Dec & \hspace{2em} \z  & $\calr$ & \\
}
\startdata
STACS J0059.6+0437 & 14.9078 & 4.6250 & 0.9 & 29.9 & 1  \\ 
STACS J0100.0+0433 & 15.0182 & 4.5600 & 0.8 & 20.2 &   \\ 
STACS J0100.3+0417 & 15.0836 & 4.2900 & 0.3 & 14.2 & 2,3  \\ 
STACS J0100.4+0427 & 15.1136 & 4.4600 & 0.6 & 37.7 &   \\ 
STACS J0100.5+0424 & 15.1287 & 4.4100 & 0.4 & 18.9 &   \\ 
STACS J0101.4+0428 & 15.3593 & 4.4800 & 0.4 & 28.6 &   \\ 
STACS J0102.6+0413 & 15.6702 & 4.2250 & 0.3 & 13.2 & 2  \\ 
STACS J0102.7+0421 & 15.6953 & 4.3600 & 0.4 & 13.1 & 2  \\ 
STACS J0102.8+0414 & 15.7053 & 4.2450 & 0.3 & 12.1 & 2,4  \\ 
STACS J0102.8+0415 & 15.7204 & 4.2550 & 0.3 & 11.4 & 5  \\ 
STACS J0102.8+0418 & 15.7204 & 4.3050 & 0.3 & 15.9 & 2  \\ 
STACS J0103.2+0419 & 15.8006 & 4.3300 & 0.4 & 37.5 &   \\ 
STACS J0103.2+0414 & 15.8156 & 4.2350 & 0.3 & 24.4 & 2,6  \\ 
STACS J0103.3+0416 & 15.8257 & 4.2800 & 0.4 & 35.7 & 2  \\ 
STACS J0104.1+0430 & 16.0464 & 4.5100 & 0.8 & 10.6 &   \\ 
STACS J0104.2+0416 & 16.0513 & 4.2750 & 0.8 & 35.0 & 6  \\ 
STACS J0104.7+0420 & 16.1968 & 4.3400 & 0.5 & 25.2 & 2 \\ 
STACS J0104.8+0422 & 16.2018 & 4.3800 & 0.5 & 20.5 & 2  \\ 
STACS J0104.8+0430 & 16.2019 & 4.5000 & 0.4 & 38.0 & 7  \\ 
STACS J0104.9+0414 & 16.2268 & 4.2400 & 0.6 & 50.6 & 2  \\ 
STACS J0105.0+0415 & 16.2669 & 4.2600 & 0.6 & 42.3 & 2,5  \\ 
STACS J0105.0+0417 & 16.2719 & 4.2850 & 0.7 & 66.9 & 2  \\ 
\enddata
\tablenotetext{a}{The Stanford Cluster Search cluster naming convention, 
    comprising the (unique) STACS acronym 
    plus the name format 'JHHMM.m+DDMM', has been registered with the
    IAU Commission 5 Task Group on Designations and conforms to their
    recommendations.}
\tablenotetext{1}{Marginal signal, visual inspection suggests it is a group.}
\tablenotetext{2}{Visual inspection indicates large angular scale spatial 
distribution and/or visually apparent
substructure, suggesting possible association with a supercluster or 
merger of clusters.  Galaxy redshift observations over a large angular 
scale indicated.  }
\tablenotetext{3}{Appearance suggests existence of a background, 
higher $z$ cluster or group close on the sky that did not generate a local
maximum in the $\lfine$ map.}
\tablenotetext{4}{Visual inspection indicates 2 bright stars and the
associated exclusion regions strongly affect the $\lfine$ map.  A cluster
near this position is consistent with the image but the center is 
poorly identified.  }
\tablenotetext{5}{May be spurious re-detection of adjacent cluster(s).}
\tablenotetext{6}{Visual inspection suggests that the center 
and/or the richness of the cluster (as indicated by the \lfine\ map peak) 
may be affected by a saturated
star and the associated excluded region.}
\tablenotetext{7}{Confirmed cluster, $z_{meas}=0.400$.  See text.  }
\end{deluxetable}

\clearpage

\begin{deluxetable}{lrrrrrrc}   
\footnotesize
\tablecaption{Confirmed cluster catalog
    \label{tab:clusterconf}}
\tablewidth{450pt}   
\tablehead{
\colhead{Name\tablenotemark{a}} &
         \multicolumn{2}{c}{Coordinates, J2000} &
         \multicolumn{1}{c}{$\calr$\tablenotemark{b}} &
         \multicolumn{1}{c}{$\langle z \rangle$} &
         \multicolumn{2}{c}{Number of redshifts} &
         \multicolumn{1}{c}{Notes:} \\
\cline{2-3} \cline{6-7} \\
      & RA\tablenotemark{a}  & Dec & & & 
       \multicolumn{1}{c}{cluster\tablenotemark{c}} & 
       \multicolumn{1}{c}{field\tablenotemark{d}} & \\
}
\startdata
STACS J0104.8+0430 &  16.2019 &  4.5000 & 40.8 & 0.400 & 6 & 10 &   \\
STACS J0823.4+0338 & 125.8654 &  3.6400 & 17.3 & 0.260 & 4 & 7 & 1 \\
STACS J1057.2$-$0340 & 164.3248 & -3.6800 & 20.3 & 0.548 & 3 & 8 & \\
\enddata
\tablenotetext{a}{Decimal degrees.}
\tablenotetext{b}{Estimated during the \lfine\ calculation.}
\tablenotetext{c}{Number in range of 2000~\kms.}
\tablenotetext{d}{Total known redshifts in field.}
\tablenotetext{1}{Candidate is within a large area above the threshold in 
   the \lfine\ map; these coordinates refer to
   a local maximum with $z_{est}=0.4$.  The center of the $z=0.26$ cluster 
   may be to
   the SE of the indicated position, while another cluster at higher
   redshift may be centered on this position, or
   one or both may be rich groups.  Further study required.}
\end{deluxetable}


\begin{thebibliography}{}
\bibitem[Abell, Corwin, \& Olowin (1989)]{abell89} Abell, G.O., 
    Corwin, H.G., \& Olowin, R.P. 1989, \apjs, 70, 1

\bibitem[Bahcall, Fan, \& Cen (1997)]{bfc97} Bahcall, N.A., 
Fan, X., \& Cen, R.\ 1997, ApJ, 485, L53

\bibitem[Bahcall \& Fan (1998)]{bf98} Bahcall, N.A., \& Fan, X.\ 
1998, \apj, 504, 1 

\bibitem[Bahcall \etal (2000)]{bcdoy00} Bahcall, N.A., Cen, R., Dav\'e, R.,
Ostriker, J.P., \& Yu, Q.\ 2000, ApJ, 541, 1

\bibitem[Balbi \etal (2000)]{b00} Balbi, A.\ \etal\ (2000), astro-ph/0005124

\bibitem[Barreiro \etal (2000)]{bhl00} Barreiro, R.B., Hobson, M.P., Lasenby,
A.N., Banday, A.J., Gorski, K.M., \& Hinshaw, G.\ 2000, \mnras, 318, 475 

\bibitem[Bartlett, Blanchard, \& Barbosa (1998)]{bbb98}
Bartlett, J.G., Blanchard, A., \& Barbosa, D.\ 1998, A\&A, 336, 425


\bibitem[Blanchard \& Bartlett (1998)]{bb98} Blanchard, A., \& 
Bartlett, J.G.\ 1998, A\&A, 332, L49

\bibitem[Blanchard \etal (1999)]{bsbd99} Blanchard, A., Sadat, R., Bartlett,
J.G., \& Le Dour, M.\ 1999, submitted to A\&A (astro-ph/9908037)

\bibitem[Borgani \etal (1999)]{borg99} Borgani, S., Rosati, P., 
Tozzi, P., \& Norman, C.\ 1999, ApJ, 517, 40

\bibitem[Bramel, Nichol, \& Pope (1999)]{bramel00} Bramel, D.A., Nichol,
R.C., \& Pope, A.C.\ 2000, \apj, 533, 601

\bibitem[Carlberg \etal (1997)]{carl97}Carlberg, R.G., Morris, S.L., 
Yee, H.K.C., \& Ellingson, E.\ 1997, ApJ, 479, L19

\bibitem[Coleman, Wu, \& Weedman (1980)]{cole80} Coleman, G.D., 
    Wu, C-C., \& Weedman, D.W. 1980, \apjs, 43, 393

\bibitem[Donahue \etal (1998)]{donahue98}Donahue, M., Voit, G.D., Gioia, I., 
    Luppino, G., Hughes, J.P., \& Stocke, J.T.\ 1998, \apj, 502, 550

\bibitem[Driver, Couch, \& Phillipps (1998)]{driver98} Driver, S.P., 
Couch, W.J., \& Phillipps, S. 1998, \mnras, 301, 369

\bibitem[Eke \etal (1996)]{ecf96} Eke, V.R., Cole, S., \& Frenk, C.S.
    1996, \mnras, 282, 263


\bibitem[Eke \etal (1998)]{ecfh98} Eke, V.R., Cole, S., Frenk, C.S., \&
Henry, J.P.\ 1998, \mnras, 298, 1145

\bibitem[Frenk \etal (1996)]{fews96}Frenk, C.S., Evrard, A.E., 
White, S.D.M., \& Summers, F.J.\ 1996, \apj, 472, 460

\bibitem[Gioia \etal (1990)]{gioia90} Gioia, I.M., Maccacaro, T., 
  Schild, R.E., Wolter, A., Stocke, J.T., Morris, S.L., and Henry, J.P.  
  1990, \apjs, 72, 567

\bibitem[Governato \etal (1999)]{g99} Governato, F., \etal\ 1999, 
\mnras, 307, 949

\bibitem[Gross \etal (1998)]{g98} Gross, M.A.K., Somerville, R.S., 
Primack, J.R., Holtzman, J., \& Klypin, A.\ 1998, \mnras, 301, 81

\bibitem[Haiman, Mohr, \& Holder (2000)]{hmh00}Haiman, Z., Mohr, J.J., \& 
Holder, G.P.\ 2000, ApJ, submitted (astro-ph/0002336)

\bibitem[Hill \etal (1998)]{hill98} Hill, G.J., Nicklas, H.E., 
MacQueen, P.J., Tejada C., Cobos Duenas, F.J., \& Mitsch, W. 1998 in Optical
Astronomical Instrumentation, ed. S. D'Odorico, (Proc. SPIE, Vol. 3355), 375


\bibitem[Hogan \& Dalcanton (2000)]{hogan00} Hogan, \& Dalcanton 2000,
    submitted to Phys. Rev. D (astro-ph/0002330)

\bibitem[Holden, \etal (1999)]{holden99} Holden, B.P., Nichol, R.C., 
   Romer, A.K., Metevier, A., Postman, M., Ulmer, M.P., \& Lubin, L.M. 
   1999, AJ, 118, 2002

\bibitem[Holden, \etal (2000)]{holden00} Holden, B.P., \etal\ 2000 AJ, 120, 23

\bibitem[Jarvis \& Tyson (1981)]{jt81} Jarvis, J.F., \& Tyson, J.A.\ 
1981, AJ, 86, 476

\bibitem[Jenkins \etal (2000)]{j00} Jenkins, A., Frenk, 
C.S., White, S.D.M., Colberg,
J.M., Cole, S., Evrard, A.E., \& Yoshida, N.\ 2000, astro-ph/0005260

\bibitem[Kawasaki \etal (1998)]{k98} Kawasaki, W., Shimasaku, K., 
Doi, M., \& Okamura, S.\ 1998, A\&AS, 130, 567

\bibitem[Kepner \etal (1999)]{kep99} Kepner, J., Fan, X., Bahcall, N.,
Gunn, J., Lupton, R., \& Xu, G. 1999, \apj, 517, 78

\bibitem[Kepner \& Kim (2000)]{kk00} Kepner, J., \& Kim, R.\ 2000,
astro-ph/0004304

\bibitem[Kinney \etal (1996)]{kinney96} Kinney, A.L., Calzetti, D., 
Bohlin, R.C., McQuade, K., 
Storchi-Bergmann, T., \& Schmitt, H.R. 1996, ApJ 467, 38

\bibitem[Kitayama \& Suto (1996)]{ks96} Kitayama, T., \& Suto, Y.\ 1996, 
ApJ, 480, 493

\bibitem[Lange \etal (2000)]{L00} Lange, A.E., \etal\ 2000, astro-ph/0005004


\bibitem[Mathiesen \& Evrard (2001)]{math01} Mathiesen, B.F. \& Evrard A.E.
2001, ApJ, in press (astro-ph/0004309)

\bibitem[Monet (1998)]{m98} Monet, D.G.\ 1998, AAS Meeting \# 194, 
Abstract 120.03

\bibitem[Navarro, Frenk, \& White (1997)]{nfw} Navarro, J.F., Frenk, 
C.S., \& White, S.D.M.\ 1997, ApJ, 490, 493


\bibitem[Nonino, \etal (1999)]{nonino99} Nonino, M. \etal\ 1999, \aaps, 137, 51



\bibitem[Oukbir \& Blanchard (1992)]{ob92} Oukbir, J., \& Blanchard,
A.\ 1992, A\&A 262, L21

\bibitem[Oukbir \& Blanchard (1997)]{ob97} Oukbir, J., \& Blanchard,
A.\ 1997, A\&A 317, 10

\bibitem[Perlmutter \etal (1999)]{perl99} Perlmutter, S., \etal\ 1999,
ApJ, 517, 565

\bibitem[Postman \etal (1996)]{post96} Postman, M., Lubin, L.M., 
    Gunn, J.E., Oke, J.B., Hoessel, J.G., Schneider, D.P., \&
    Christensen, J.A. 1996, \aj, 111, 615

\bibitem[Postman \etal (1998)]{deeprange} Postman, M., Lauer, T.R., Szapudi,
I., \& Oegerle, W.\ 1998, ApJ, 506, 33

\bibitem[Press \& Schechter (1974)]{ps} Press, W.H., \& Schechter, P.\ 
1974, ApJ, 187, 425 

\bibitem[Press \etal (1986)]{press86} Press, W.H., Flannery, B.P., 
    Teukolsky, S.A., \& Vetterling, W.T. 1986, Numerical Recipes 
    (Cambridge: Cambridge University Press)

\bibitem[HET; Ramsey \etal (1998)]{het98} Ramsey L.W., et al. 1998, Proc.
SPIE, 3352, 34

\bibitem[Reichart \etal (1999)]{R99} Reichart, D.E.\ \etal\ 1999, ApJ, 518, 521

\bibitem[Romer \etal (1999)]{rvlm99} Romer, A.K., Viana, P.T.P., Liddle, A.R.,
\& Mann, R.G.\ 1999, astro-ph/9911499 

\bibitem[Sadat, Blanchard, \& Oukbir (1998)]{sbo98} Sadat, R., Blanchard, A.,
\& Oukbir, J.\ 1998, A\&A, 329, 21

\bibitem[Schechter (1976)]{s76} Schechter, P.\ 1976, ApJ, 203, 297

\bibitem[Schlegel, Finkbeiner, \& Davis (1998)]{schlegel98} Schlegel, D.J., 
    Finkbeiner, D.P., \& Davis, M. 1998, \apj, 500, 525

\bibitem[Schuecker \& Boehringer(1998)]{sb98} Schuecker, P., \& 
Boehringer, H.\ 1998, A\&A, 339, 315


\bibitem[Sheth \& Tormen (1999)]{st99} Sheth, R.K., \& Tormen, G.\ 1999, 
MNRAS, 308, 119

\bibitem[Sheth, Mo, \& Tormen (1999)]{smt99} Sheth, R.K., Mo, H.J.,
\& Tormen, G.\ 1999, astro-ph/9907024

\bibitem[Spergel \& Steinhardt (2000)]{sperg00} Spergel, D.N., \& 
    Steinhardt, P.J. 2000, \prl, 84, 3760

\bibitem[Stocke \etal (1991)]{stocke91} Stocke, J.T., Morris, S.L., 
    Gioia, I.M., Maccacaro, T., Schild, R., Wolter, A., Flemming, T.A., \&
    Henry, J.P. 1991, \apjs, 76, 813

\bibitem[Tran \etal (1999)]{tran99} Tran, K-V. H., Kelson, D.D., 
    van Dokkum, P., Franx, M., Illingworth, G.D., \& Magee, D. 1999, 
    \apj, 522, 39

\bibitem[van Haarlem, Frenk, \& White (1997)]{vhfw97} van Haarlem, M.P., Frenk,
C.S., \& White, S.D.M.\ 1997, MNRAS, 287, 817

\bibitem[Viana \& Liddle (1999a)]{vl99a} Viana, P.T.P., \& Liddle, A.R.\
(1999a), \mnras, 303, 535

\bibitem[Viana \& Liddle (1999b)]{vl99b} Viana, P.T.P., \& Liddle, A.R.\
(1999b), astro-ph/9902245

\bibitem[White, Efstathiou, \& Frenk (1993)]{wef93} White, S.D.M., 
Efstathiou, G., \& Frenk, C.S.\ 1993, MNRAS, 262, 1023

\bibitem[Willick (1999)]{w99} Willick, J.A. 1999, ApJ, 516, 47

\bibitem[Willick (2000)]{w00} Willick, J.A.\ 2000, ApJ, 530, 80

\end{thebibliography}
\end{document}